\documentclass[showpacs,preprintnumbers,nofootinbib,amsmath,amssymb]{revtex4}

\usepackage{graphicx,color}
\usepackage{url}
\usepackage{latexsym}
\usepackage{multirow}
\usepackage{CJK}
\pdfoutput=1
\begin{document}
\input{epsf.sty}

\def\affilmrk#1{$^{#1}$}
\def\affilmk#1#2{$^{#1}$#2;}
\def\be{\begin{equation}}
\def\ee{\end{equation}}
\def\bea{\begin{eqnarray}}
\def\eea{\end{eqnarray}}


\centerline{\bf \Large Introduction of the CDEX experiment}

\vspace{0.5cm}


\vspace{1cm}

\begin{center}
Ke-Jun KANG$^1$,Jian-Ping CHENG$^1$,Jin LI$^1$, Yuan-Jing LI$^1$, Qian YUE$^1$, Yang BAI$^3$,Yong BI$^5$, Jian-Ping CHANG$^1$, Nan CHEN$^1$, Ning Chen$^1$, Qing-Hao CHEN$^1$, Yun-Hua CHEN$^6$, Zhi DENG$^1$,Qiang DU$^1$, Hui GONG$^1$, Xi-Qing HAO$^1$, Hong-Jian HE$^1$, Qing-Ju HE$^1$, Xin-Hui HU$^3$, Han-Xiong HUANG$^2$, Hao JIANG$^1$, Jian-Min LI$^1$, Xia LI$^2$, Xin-Ying LI$^3$, Xue-Qian LI$^{3,1}$, Yu-Lan LI$^1$, Shu-Kui LIU$^5$, Ya-Bin LIU$^1$, Lan-Chun L\"{u}$^1$, Hao MA$^1$, Jian-Qiang QIN$^1$, Jie REN$^2$, Jing Ren$^1$, Xi-Chao RUAN$^2$, Man-Bin SHEN$^6$, Jian SU$^1$, Chang-Jian TANG$^5$, Zhen-Yu TANG$^5$, Ji-Min WANG$^6$, Qing WANG$^{1,}$\footnote{corresponding
author.\\ E-mail addresses: lixq@nankai.edu.cn(Xue-Qian Li),wangq@mail.tsinghua.edu.cn(Qing WANG)}, Xu-Feng Wang$^1$, Shi-Yong WU$^6$, Yu-Cheng WU$^1$, Zhong-Zhi Xianyu$^1$, Hao-Yang XING$^5$, Xun-Jie Xu$^1$, Yin XU$^3$, Tao XUE$^1$, Li-Tao YANG$^1$, Nan YI$^1$, Hao YU$^1$, Chun-Xu YU$^3$, Xiong-Hui ZENG$^6$, Zhi ZENG$^1$, Lan ZHANG$^4$, Guang-Hua ZHANG$^6$, Ming-Gang ZHAO$^3$, Su-Ning ZHONG$^3$, Jin ZHOU$^6$, Zu-Ying ZHOU$^2$, Jing-Jun ZHU$^5$, Wei-Bin ZHU$^4$, Xue-Zhou ZHU$^1$, Zhong-Hua ZHU$^6$

\vspace*{0.7cm}
(CDEX Collaboration)

\vspace*{0.7cm}
$^1$Tsinghua University, Beijing, China£¬100084\\
$^2$Institute Of Atomic Energy, Beijing, China, 102413\\
$^3$Nankai University, Tianjin, China, 300071\\
$^4$NUCTECH company, Beijing, China, 100084\\
$^5$Sichuan University, Chengdu, China, 610065\\
$^6$Yalongjiang Hydropower Development Company, Chengdu, China, 627450
\end{center}

\vspace{0.8cm}
\begin{center}
{\bf Abstract}\vspace*{0.5cm}~

\begin{minipage}{17cm}
~~~~Weakly Interacting Massive Particles (WIMPs) are the candidates
of dark matter in our universe. Up to now any direct interaction of
WIMP with nuclei has not been observed yet. The exclusion limits of
the spin-independent cross section of WIMP-nucleon which have been
experimentally obtained is about $\mathrm{10^{-7}pb}$ at high mass
region and only $\mathrm{10^{-5}pb}$ at low mass region. China
Jin-Ping underground laboratory CJPL is the deepest underground lab
in the world and provides a very promising environment for direct
observation of dark matter. The China Dark Matter Experiment (CDEX)
experiment is going to directly detect the WIMP flux with high
sensitivity in the low mass region. Both CJPL and CDEX have achieved
a remarkable progress in recent two years. The CDEX employs a
point-contact germanium semi-conductor detector PCGe whose detection
threshold is less than 300 eV. We report the measurement results of
Muon flux, monitoring of radioactivity and Radon concentration
carried out in CJPL, as well describe the structure and performance
of the 1 kg PCGe detector CDEX-1 and 10kg detector array CDEX-10
including the detectors, electronics, shielding and cooling systems.
Finally we discuss the physics goals of the CDEX-1, CDEX-10 and the
future CDEX-1T detectors.

\end{minipage}
\end{center}
\vspace{1cm}

\section{Introduction}

Discovery of dark matter undoubtedly was one of the greatest
scientific events of the 20th  century, then directly searching for
dark matter and identifying it will be the most important and
challengeable task of this century.

As a matter of fact, the conjecture about  existence of dark matter
was proposed quite a long time ago in 1933 by Zwicky \cite{zwicky}
to explain the anomalously large velocity  of test stars near the
Coma star astronomically observed at that time. The astronomical
observation shows that the rotational curves of the test stars in
the galaxy did not obey the gravitational law if only the luminous
matter which clusters at the center of the galaxy existed. Namely,
the velocities of test stars were supposed to be inversely
proportional to square roots of their distances from the center of
the galaxy, instead, the rotational curve turns flat. It implies
that there must be some unseen matter in the galaxy, i.e. the dark
matter. Moreover,  hints about existence of dark matter also appear
when a collision of two clusters was observed. It was observed that
the center of mass of each cluster does not coincide with its center
of the luminous matter after a collision between two clusters. It is
explained as that two clusters both of which are composed of dark
matter and luminous matter, collide, after collisions, the dark
components penetrate each other because they do not participate in
electromagnetic interaction (EM) neither strong interaction, but the
luminous fractions of the two clusters interact with each other via
EM interaction, so remain near the collision region while the dark
parts left.

Moreover, all astronomical observations confirm that our universe is
approximately flat, i.e. the total $\Omega$  defined as
$\rho/\rho_c$ where $\rho_c$ is the critical density and $\rho$ is
the matter density in our universe, is close to unity. However,
observation also indicates that the fraction of luminous baryonic
density $\Omega_m$ is less than 4\%. By fitting the observational
data, it is confirmed that over 96\% of matter is dark. Further
analysis indicates that the dark matter may take a fraction of 24\%
while the dark energy occupies the rest over 72\%. The dark energy
is the most mysterious subject for which so far our understanding of
the universe is not enough to give a reasonable answer even though
there are many plausible models. By contrary, the dark matter may
have a particle correspondence.

Commonly accepted point of view~\cite{v.trimble} is that the main
fraction of the dark matter in our universe is the cold dark matter,
i.e. the weakly-interacting-massive-particles (WIMPs). The criteria
of being a dark matter candidate are that the particle does not
participate in EM interaction (so it is dark!), namely it must be
neutral £¨both electric and color) and does not possess inner
structure, otherwise it may have intrinsic anomalous magnetic
moments and interact with EM field. Thus generally, it is the so
called elementary particles. The most favorable candidate of the
WIMPs is the neutralino~\cite{em}, even though one cannot exclude
other possibilities at the present~\cite{lars bergstrom,j.l.feng}.
For example, He et al. \cite{He} proposed that a scalar particle
darkon can be a possible dark matter candidate which interacts with
detector matter via exchange of Higgs boson. Besides, the
technicolor meson \cite{beylyaev}, asymmetric mirror dark matter
\cite{an:2010kc}, new particles in the little Higgs model
\cite{gilloz} as well as many other candidates have also been
predicted by various models as the dark matter candidates. It is
interesting that, thanks to the LHC which has already successfully
been operating, one may search for such beyond standard model
particles at accelerators.

One of the main goals of the detection of the dark matter  is to
help finding and identifying the dark matter particle(s).

To realize the task, one should design sort of experiments.
Detection of dark matter is by no means a trivial job. It can be
classified into the direct and indirect detections. For the first
one, we should set an underground detector and try to catch the dark
matter flux from outside space. In this scheme it is supposed that
the dark matter flux comes from outer space, such as the halo of
galaxy or even the sun and the dark matter particles (WIMPs) weakly
interact with the standard model (SM) matter (mainly quarks).
However, we really do not have a solid knowledge to guarantee that
the dark matter particles indeed weakly interact with the regular
matter as described by the gauge theory. It is just like that a
black cat is confined in a dark room where there is no light, and we
are supposed to  catch it. Even though the chance of catching it is
slim, we still have probability to get it. But if it does not exist
in the room, no matter how hard we strive, we can never
succeed\footnote{This is a story presented by Prof. T. Han at a
Jin-Ping conference on dark matter detection, Sichuan, China,
2011.}. Therefore, generally our project of directly detecting the
dark matter flux is based on such hypothesis that it does interact
with the detector matter via weak interaction.

Another line is indirect detection of dark matter. In that scheme,
it is hypothesized that the dark matter particle may decay or
annihilate into SM particles via collisions among dark matter
particles. This proposal is to explain the peculiar phenomena of
$e^+$ excess \cite{excess} and the cosmic-ray energy spectrum
{\cite{tibet} observed by the earth laboratories and satellites.

The China Dark Matter Experiment (CDEX) is designed to directly
detect the dark matter with the highly pure germanium detector, thus
later in this work we will concentrate ourselves on the discussion
of direct search for dark matter. Unfortunately, the kinetic energy
of the dark matter particle ${1\over 2}m\beta^2$ is rather low,
generally the velocity of the WIMPs is $200\sim 1000$ km/s and the
value of $\beta=v/c$ is about $10^{-3}\sim 10^{-2}$, thus for a WIMP
particle of 50 GeV, its kinetic energy is only a few tens of keV,
which is too small to cause an inelastic transition for the nucleus
(not exactly, see below discussions). When the WIMPs hit on the
nucleus in the detector material, the impact makes the nucleus to
recoil and then the atom would be ionized. When the nucleus is
recombined with the electron clouds, photons may be radiated
accompanied by the electric and thermal signals which can be
detected by sensitive detectors. Each detector may be designed to be
sensitive to one or a few of the signals and then record them to be
analyzed off-line.

Recently, several groups have reported limited  success for
detecting dark matter flux. The DAMA \cite{dama} reported their
observation of the annual modulation signal and then CoGeNT
\cite{cogent} observed several cosmogenic peaks and their results
are consistent with the DAMA's. The CDMS group reported to observe a
few events of dark matter flux \cite{Brink} and recently CRESST-II
published their new observation results\cite{cresst} and numerous
underground laboratories are working hard to search for the signals
of dark matter. With its special advantages, such as the thickest
rock covering which can shield out most of the cosmic rays and
convenient transportation and comfortable working and living
condition, the Jin-Ping underground laboratory would provide an
ideal circumstance for the dark matter detection and the new CDEX
collaboration has joined the club for direct dark matter search. The
detailed descriptions about the CDEX project will be given in the
following sections.

The detection is carried out via collisions  between WIMPs and
nuclei. To extract any information about the fundamental interaction
(SUSY, technicolor, darkon or even little Higgs etc.) from the data,
much theoretical work and careful analysis must be done. For the
theoretical preparation, one has to divide the whole process into
three stages. The first is to consider the elementary scattering
between WIMPs and quarks or gluons inside the nucleons, from there
we need to derive an effective Hamiltonian describing elastic
scattering between DM particle and nucleon, then the subsequent
stage is the nuclear stage. Since the kinetic energy of WIMPs is
rather low and generally the elastic transition dominates, the
energy absorbed by the nucleons would eventually pass to the whole
nucleus which recoils as described above. But this allegation is not
absolutely true. Indeed the gap between radial excitations of
nucleus (the principal quantum numbers of valence nucleons change)
and its ground state is rather large comparing with the available
collision kinetic energy. However, if the ${\bf L}\cdot {\bf S}$
coupling is taken into account, the energy levels which were
degenerate would be split and the resultant gaps between the split
levels could be small and comparable with the kinetic energy of the
dark matter particles, thus the collision between the dark matter
particle and nucleon can induce an inelastic transition and radiate
photons. But this is not our concern for the CDEX project.

The highly pure germanium detector is designed to be sensitive to
low energy dark matter flux.

The interaction can also be categorized into  spin-independent and
spin-dependent ones. So far, the measurements on the
spin-independent WIMP-nucleon cross section can reach an accuracy to
$10^{-44}$ cm$^2$, but for the spin-dependent case, it is only about
$10^{-39}$ cm$^2$. It is believed that beyond $10^{-46}$ cm$^2$, the
cosmic neutrino would compose unavoidable background and the
measurements do not make any sense after all.

\section{A Brief Review of Theoretical framework for detection of dark matter flux}

\subsection{Kinematics}

We are dealing with collisions between WIMPs  and nucleus, let us
first present the necessary expressions which are related to the
detection.

The recoil energy of the nucleus is \cite{an:2010kc}
\begin{equation}
Q={1\over 2}mv^2={2mM\over (m+M)^2}(1-\cos\theta_{cm})={\mu_{red}^2v^2(1-\cos\theta_{cm})},
\end{equation}
where m, M stand for the masses of the WIMP  and  nucleus, $v$ is
the absolute value of the velocity of the WIMP in the laboratory
frame, $\mu_{red}$ is the reduced mass of the WIMP and nucleus and
$\theta_{cm}$ is the scattering angle in the center of mass frame of
the WIMP and nucleus. The recoil momentum is
\begin{equation}
|{\bf q}|^2=2\mu_{red}^2 v^2(1-\cos\theta_{cm}).
\end{equation}
The recoil rate per unit detector mass is
\begin{equation}
R={n\sigma v\over M_D}={\rho\sigma v\over mM_D},
\end{equation}
where $\rho$ is the local dark matter density  which is 0.3
GeV/cm$^3$ in the standard halo model. The differential rate is
\cite{jungman:1995}
\begin{equation}
{dR\over dQ}=2M_D{dR\over d|{\bf q}|^2}={2\rho\over m}\int vf(v) {d\sigma\over d|{\bf q}|^2}(|{\bf q}|^2=0)d^3 v,
\end{equation}
where $f(v)$ is the velocity distribution function of WIMP and
\begin{equation}
{d\sigma\over d|{\bf q}|^2}(|{\bf q}|^2=0)={\sigma_0\over 4\mu_{red}^2 v^2},
\end{equation}
with $\sigma_0$ being the scattering cross section at the zero
momentum transfer limit. The function $f(v)$ of DM in the galactic
halo is assumed to be in the Gaussian form \cite{an:2010kc} as
\begin{equation}
f({\bf u})=f({\bf v}+{\bf v_e})={1\over (\pi v_0^2)^{3/2}}{e^{-{\bf u}^2\over v_0^2}},
\end{equation}
where $v_0\sim 270$ km/s, ${\bf v}$ is the velocity of  the DM with
respect to the detector and ${\bf v_e}$ is the velocity of the earth
with $v_e=v_{\odot} +14.4 \cos[2\pi(t-t_0)/T]$, and $t_0=152$ days,
$T=1$ year which reflects the annual modulation effects. It is noted
that here $f({\bf u})$ is exactly $f(v)$ which we used in above
expressions.  The average velocity of the dark matter is 200 km/s
$\sim$ 600 km/s, the kinetic energy  ${1\over 2}m({v\over c})^2$ is
as small as 20$\sim$ 200 keV as m=100 GeV. That is the maximum
energy which can be transferred to the nucleus and cause its recoil.
If the dark matter particle is as light as 10 GeV, the kinetic
energy is only at the order of a few hundreds of eV. The small
available energy results in a small recoil energy of the nucleus and
demands a high sensitivity of our detector at low energy ranges.
That raises a serious requirement for the detector design and it is
the goal of the CDEX. It is worth noticing that for very low mass DM
candidates which are predicted by some special models, or very low
recoil energy, the contribution of DM-electron scattering is not
negligible, moreover, the recoil of electron may compose the main
signal\cite{essig}.
\subsection{Cross section and amplitude}

The calculation of the cross section of the collision between the DM
particle and nucleus can be divided into three stages: the first one
is calculating the amplitudes of the elementary collision with
quarks, the second stage is calculating the amplitudes for collision
with the nucleon (proton and neutron), then the third stage is
calculating the cross section for collision with the whole nucleus
if one can assume that the nucleus is free. But in fact it is bound
on lattice of crystals, thus when the small temperature effects are
concerned a fourth stage calculation is required.

The WIMP particle interacts with quarks or gluons  inside the
nucleon~\cite{ressell,griest,shan:2011ct,shan:2011jz}. Gluons in the
hadron do not participate in the weak interaction at the leading
order, so that the fundamental processes concern only the
interaction between DM particle and quarks. At the second stage, one
needs to write up the effective Hamiltonian for the interaction
between the DM particle and nucleon from the fundamental
interaction. Because it is an elastic process, the nucleon remains
unchanged after the collision and the total exchanged energy is
transformed into the kinetic energy of the nucleon. Obtaining the
effective Hamiltonian for the DM particle-nucleon interaction is by
no means trivial. We will discuss the procedure more explicitly when
we talk about the SI and SD processes in next subsections. Generally
the collisions between dark matter particles with nucleus are very
low energy processes, nucleus cannot be excited to higher radial
states, thus the processes are elastic. Indeed, there might be
orbital excitations and the collision would be inelastic, in general
such processes only cause very small observable effects. Later the
excited nucleus radiates a photon and returns to the ground state,
but for the germanium detector which cannot detect photon radiation,
such effects do not need to be considered. We only concentrate on
the nucleus recoil effects. Definitely, the goal of the whole
research is to identify the DM particle(s) and discover the
interaction which may be new physics beyond the standard model. But
for this purpose, we need to study the observable effects and learn
how to extract useful information about the fundamental processes.

The cross section between WIMP and nucleus is
$$\sigma={1\over mMv}{1\over (2s_1+1)(2s_2+1)}\sum\int {d^3p'_1\over
(2\pi)^3}{1 \over 2E'_1}{d^3p'_2\over (2\pi)^3}{1 \over
2E'_2}|M|^2(2\pi)^4\delta (p_1+p_2-p'_1-p'_2),$$ where $v$ is the
velocity of the DM particle while the nucleus is assumed to be at
rest before collision, $p_1,p_2,p'_1,p'_2$ are the momenta of the DM
particle and nucleus in the initial and final states respectively,
$s_1,s_2$ are the spins of the DM particle and nucleus, the sum is
over all the polarizations of the outgoing DM particle and nucleus,
$M$ is the collision amplitude which is what we need to calculate.

How to get the effective coupling of $DM-N\bar N$  depends on the
concrete model adopted in the
calculation~\cite{barger:2008qd,tzeng:1996ve,shan:2011ct,shan:2011jz}.
For example, the exchanged meson between the DM particle and nucleon
could be the SM $Z^0$ or Higgs boson \cite{cheng}, whereas in the
models beyond the standard model the exchanged agents may be $Z'$ or
others.

\subsubsection{Amplitude}

The elementary process is an elastic scattering between the DM
particle and quarks. For example in the standard model (SM) the
exchanged boson between DM particle and quark (u,d,s) is $Z^0$ or
Higgs boson whereas in the new physics beyond the SM, it may be $Z'$
etc. The Lagrangian which is determined by the pre-postulated models
of the dark matter and the effective interaction is written as
\begin{equation}
{\cal L}=l_{\Gamma} \bar q \Gamma q \;\;\; (q=u.d,s),
\end{equation}
where $\Gamma$ is a combination of the $\gamma$ matrices
($S,\;\gamma_\mu,\; i\sigma_{\mu\nu},\; \gamma_\mu\gamma_5,\;
\gamma_5$) and transfer momentum $q$, $l_{\Gamma}$ is the
corresponding DM current. It is noted that only the scalar $S$ and
axial vector $\gamma^\mu\gamma_5$ do not suffer a suppression of
smaller transfer momenta. The former corresponds to the SI process
which will be discussed in next subsection and the later one is
proportional to the quark spin under the non-relativistic
approximation and results in the spin-dependent (SD) cross section.

According to the general principle of the quantum field theory, the
amplitude of the elastic scattering between the dark matter particle
with a single nucleon should be written in the momentum space
because during the process the total energy and the three-momentum
are conserved. Thus the amplitude is
\begin{equation}\label{amp}
M=l_{\mu}\bar u({\bf p}',s'_z)J^{\mu} u({\bf p},s_z),
\end{equation}
where $l_{\mu}$ is the DM current, ${\bf p}\;,{\bf p}',\,s_z,\,s'_z$
are the three-momenta and spin projections of the initial and final
states of the nucleon and $J^{\mu}$ is the corresponding nucleon
current of the effective interaction (for such as the darkon models,
$J^{\mu}$ would be replaced by a scalar, for more complicated
models, it might even be tensors). If the nucleon is free, the
amplitude can be easily calculated as depicted in any textbook of
quantum field theory. However the nucleon is not free but bound in
the nucleus, thus the wavefunction which describes the state of the
nucleon is a solution of the Schr\"odinger equation with very
complicated nuclear potential, such as the Paris potential, written
in the coordinate space. Thus, one needs to derive the formula in
terms of the given wavefunctions via a Fourier transformation.
\begin{eqnarray}
M &=& l_{\mu}\bar u({\bf p}',s'_z)J^{\mu} u({\bf p},s_z) \nonumber \\
&=& l_{\mu}{1\over (2\pi)^3}\int d^3x' e^{-i{\bf p}'\cdot {\bf x}'}\bar
u({\bf x}',s_z') J^{\mu} {1\over (2\pi)^3}\int d^3x e^{i{\bf p}\cdot
{\bf x}}u({\bf x},s_z)\times (2\pi)^3\delta({\bf x}'-{\bf x}) \nonumber\\
&=& l_{\mu}{1\over (2\pi)^3}\int e^{-i({\bf p}'-{\bf p})\cdot {\bf
x}}\bar u({\bf x},s_z') J^{\mu} u({\bf x},s_z)\nonumber\\
&=& l_{\mu}{1\over (2\pi)^3}\int d^3x e^{-i{\bf q}\cdot {\bf x}}\bar
u({\bf x},s_z') J^{\mu} u({\bf x},s_z),
\end{eqnarray}
where $\bar u({\bf x},s_z), \; u({\bf x},s'_z)$ are the wave
functions of the nucleon in the coordinate space. The delta function
$(2\pi)^3\delta({\bf x}'-{\bf x})$ is introduced because the
interaction occurs at the same point (i.e. the propagator is reduced
as ${1\over q^2-M^2}\approx {-1\over M^2}$  where $M$ is the mass of
the medium boson which is heavy and $M^2$ is much larger than
$q^2$).  Now let us step forward to discuss the scattering amplitude
between the DM particle and nucleus. We re-interpret the wave
function $u(x,t)$ in the above expression as the corresponding field
operator in the second quantization scheme. The nucleon field
operator $\Psi_N({\bf r}, t)$ can be written  as
\begin{equation}
\Psi_N({\bf r},t)=\sum_{\alpha}[a_{\alpha}u_{\alpha}({\bf r})e^{-i
E_{\alpha} t}+b_{\alpha}^{\dag}\nu_{\alpha}({\bf r})e^{i
E_{\alpha}t}].
\end{equation}
where $a_{\alpha}$ and $b_{\alpha}^{+}$ are annihilation and
creation operators of baryon and anti-baryon. The probability
operator of the nucleon density at ${\bf r}$ is
\begin{equation}\label{nucleon}
\Psi_N^{\dagger}({\bf r},t)\Psi_N({\bf r},t).
\end{equation}

The nuclear ground state can be expressed as $$|\Psi_{A}>=e^{i{\bf
P}\cdot {\bf X}}(\Pi_{i=1}^A a_i ^{\dag}  )|0>,$$ where $|0>$ is the
vacuum, A is the number of nucleons in the nucleus and the product
corresponds to creating A nucleons (protons and neutrons) which are
the energy eigenstates. The phase factor $e^{i{\bf P}\cdot {\bf X}}$
corresponds to the degree of freedom of the mass center of the
nucleus, where ${\bf P}$ is the total momentum of the nucleus and
${\bf X}$ is the coordinate of the mass center of the nucleus. In
the center of mass frame, ${\bf P}=0$, the phase factor does not
show up. Indeed, the phase factor only corresponds to a free
nucleus, instead, if one further considers that the nucleus is bound
on the lattice of crystal such as germanium, the phase factor would
be replaced by a complicated function. Unless we discuss the thermal
effect such as for the CDMS detector, the nucleus can be treated as
a free particle without causing much errors.  The thermal effect can
be calculated in terms of the phonon theory. However, in most cases,
the nucleus can be treated as a free particle, and a simple phase
factor sufficiently describes this degree of freedom.

Sandwiching  the probability operator Eq. (\ref{amp}) between the
nuclear ground state, we have
\begin{equation}\label{total}
\int d^3x e^{i{\bf q}\cdot{\bf x}}<0|e^{-i{\bf P}'\cdot {\bf X}}(\Pi_{i=1}^A a_i)\sum_{\alpha\beta} a_{\alpha}^{\dagger}\bar u_{\alpha}({\bf x})e^{i\omega_{\alpha}t}
(J^{\mu})a^{\beta} u_{\beta}({\bf x}) e^{-i\omega_{\beta}t}(\Pi_{i=1}^A a_i^{\dagger})e^{i{\bf P}\cdot {\bf X}}|0>.
\end{equation}

Now, let us turn to the laboratory reference frame where the recoil
of the nucleus must be clearly described. By doing so, we need to
consider the phase factor $e^{i{\bf P}\cdot {\bf X}}$ in Eq.(14). As
aforementioned the scattering between the DM particle and nucleus is
elastic, thus any single nucleon cannot transit from its original
state to an excited state, but transfers its kinetic energy gained
from the collision with the DM particle to the whole nucleus.  The
practical process of the energy transfer is via interaction among
all the nucleons in the nucleus, so must be very complicated.
Fortunately, we do not need to know all details of the process.
Since the inner state of the nucleus does not change after the
collision, the wavefunction of the nucleus can only vary by a phase
factor. We consider that the time duration of the energy transfer
process is very short compared with our measurement, we may use the
"impulse" approximation, i.e. the nucleus does not have enough time
to move. The phase factor obviously does not depend on the
coordinates of any individual nucleon, so can only be related to the
coordinate of the mass center of the nucleus ${\bf X}$. Including
this phase factor, we would rewrite the expression (14) as
\begin{eqnarray}
&&\hspace{-0.5cm} \int d^3X <0|e^{-i{\bf P}'\cdot {\bf X}}
e^{-i{\bf q}\cdot {\bf X}}\int d^3x(\Pi_{i=1}^A a_i)\sum_{\alpha\beta} a_{\alpha}\bar u_{\alpha}e^{i\omega_{\alpha}t}(J^{\mu})
a^{\beta\dagger} u_{\beta} e^{-i\omega_{\beta}t}e^{-i({\bf p}_{\alpha}\cdot {\bf x}_{\alpha}-{\bf p'}_{\beta}\cdot {\bf x'}_{\beta})}
(\Pi_{i=1}^A a_i^{\dagger})e^{i{\bf P}\cdot {\bf X}}|0>\nonumber\\
&=& (2\pi)^3\delta({\bf P}-{\bf P}'-{\bf q})\int d^3x<0|(\Pi_{i=1}^A
a_i)\sum_{\alpha\beta} a_{\alpha}\bar
u_{\alpha}e^{i\omega_{\alpha}t}(J^{\mu}) a^{\beta\dagger} u_{\beta}
e^{-i\omega_{\beta}t}(\Pi_{i=1}^A a_i^{\dagger})e^{-i({\bf
q}\cdot {\bf
x})}|0>,
\end{eqnarray}
where the $\delta-$function explicitly embodies the momentum
conservation. For simplicity, below, we will not show this factor
anymore, as well understood. It is shown that for elastic scattering
$\alpha=\beta$, so the factor
$e^{-i\omega_{\beta}t}e^{i\omega_{\alpha}t}=1$, but if, as
aforementioned, there exists orbital excitation, this factor is no
longer 1, but in general cases, it is indeed very close to unity.

There are two types of cross sections : the SI$-$ spin-independent and SD$-$ spin-dependent ones.

\subsubsection{The SI cross section}

The fundamental interaction between the DM particle and quarks is
gained based on models, various models would result in different
Hamiltonian. Nevertheless,  from the fundamental interaction between
the DM particle and quarks to the level of nucleons, the procedure
was explicitly demonstrated in Refs.\cite{cheng,he,ellis}, and the
readers are suggested to refer to those enlightening works.

As noted, if the interaction between the DM particle and quarks is
spin-independent, for example the darkon case \cite{he}, the
particle interaction and the nuclear effect can be factorized and an
enhancement factor proportional to $A$ appears.

The observation rates for spin-independent scattering can be written as~\cite{jungman:1995}
\begin{eqnarray}\label{event}
\frac{d\sigma}{d|{\bf{q}}|^{2}}&=&G_{F}^{2}\frac{C}{v^2}F^{2}(|{\bf{q}}|)
=\frac{\sigma_{0}}{4m_{r}^{2}v^{2}}F^{2}(|{\bf{q}}|),
\end{eqnarray}
where $G_F$ is the universal Fermi coupling  constant, $\sigma_0$ is
the cross section at zero-recoil, $m_r$ is the reduced mass of the
WIMP and nucleus, finally $F(|{\bf{q}}|)$ is the nuclear form
factor. Generally speaking, the mass density distribution of the
nucleus is proportional to the charge density or the nucleon number
density, hence the form factor can also be accounted from the
nucleon number density in the nucleus, i.e. the nuclear density. The
most general form is given in Eq. (14), but for the SI cross
section, the recoil effect can be attributed into a simple form
factor. The form factor is the Fourier transformation of the nuclear
density as
\begin{eqnarray}\label{fourier transform}
F(q)&=&\frac{1}{A}\int \rho(r)e^{-i{\bf{q}}\cdot
{\bf{r}}}d^3r \nonumber\\
&=&\frac{4\pi}{A}\int \frac{r}{q} \rho(r)\sin{(qr)}dr,
\end{eqnarray}
where $\rho(r)$ is the nuclear density. Here we assume that  the
nucleus is spherically symmetric and it is only a function of $r$.
Obviously, unless one can more accurately determine $F(q)$,
extraction of useful information from data is impossible.

There are several ansatz for determining the form factor by assuming
typical $\rho(r)$ functions~\cite{helm,sogmodel,fbmodel,jdl}.

The form factor $F(q)$ is required to obey $$F(0)=1.$$

The numerical results show \cite{rel} that the form factors
determined by the various models for the nuclear density do not
deviate much from each other, therefore for  not very high accuracy
of measurement, one can use either of the models.

Generally, the SI cross section can be written as
\begin{equation}
\sigma_0^{SI}={4\over\pi}m_r^2[Zf_p+(A-Z)f_n]^2,
\end{equation}
and eventually we have
\begin{equation}
{dR\over dQ^2}={\rho_0\sigma_0\over 2m_{\chi}m_r^2}F(Q)^2\int_{v_{min}}^{v_{max}}{f(v)\over v}dv,
\end{equation}
where $Q^2=-q^2$ in the time-like form. Here $f_p$ and  $f_n$ are
related to the proton and neutron respectively which can be derived
from the given effective couplings between DM particle and nucleon
discussed above. To a good approximation
\begin{equation}
f_n\approx f_p,
\end{equation}
thus $\sigma_0^{SI}\propto A^2$ is a  large enhancement factor,
especially for heavy nuclei. Therefore, thanks to the enhancement
factor, the total spin-independent cross section could be much
larger than the cross section for DM particle scattering with single
nucleons. That is why, the present data can reach an accuracy of
$10^{-44}$ cm$^2$ for SI cross section between DM and nucleon.

\subsubsection{The SD cross section}

For the elastic scattering with small momentum transfer, the
contribution of the axial vector is dominant
\begin{equation} g^2 l^{\mu}\bar
q\gamma_{\mu}\gamma_5 q\approx g^2 {\rm\bf l}\cdot
\bar q{\bf s} q,\;\;\;\; (q=u,d.s),
\end{equation}
where $g$ is the coupling in the concerned theory and $l^{\mu}$ is
the leptonic current (could be vector and/or axial vector). Summing
over the contributions of all partonic spin projections, one would
step to the interaction between the DM particle and nucleon. Thus
the effective current for nucleon
is~\cite{ressell:1993qm,ressell:1997kx}
\begin{equation}
<p(n)|\sum_{u,d,s}g\bar q s_z q|p(n)>=\sum_{q=u,d,s}A^{p(n)}_q\Delta_q,
\end{equation}
where  $\Delta q\equiv s_q(\uparrow)-s_q(\downarrow)+\bar
s_q(\uparrow)-\bar s_q(\downarrow)$, $A^p_q$, $A^n_q$ can be
obtained from the fundamental Hamiltonian\cite{ressell}. By the
data, we have
$$\Delta_u^{(p)}=0.78\pm 0.02,\;\;\Delta_d^{(p)}=-0.48\pm 0.02,\;\;\Delta_s^{(p)}=-0.15\pm
0.02,$$ and $\Delta_u^{(n)}=\Delta_d^{(p)};\;\;\Delta_d^{(n)}=\Delta_u^{(p)};\;\;\Delta_s^{(n)}=\Delta_s^{(p)}.$

Calculating the SD cross section is  much more complicated than that
for SI cross section. For the SD process, the nuclear effect cannot
be factorized out, but entangled with the fundamental sub-processes.

The unsuppressed effective nucleon current is~\cite{engel,engel:1991wq}
\begin{equation}\label{induce}
<p,s|{\cal J}_5^{\mu}(x)|p',s'>=\bar u_N(p,s){1\over 2}[(a_0+a_1\tau_3)\gamma^\mu\gamma_5+
(b_0+b_1\tau_3)q^{\mu}\gamma_5]u_N(p',s') e^{i q\cdot x}
\end{equation}
where $q=p-p'$ and $a_0, a_1$  are given as
\begin{eqnarray}
a_{p} &=& a_0+a_1=\sum_{q=u,d,s}A^p_q\Delta_q;\\
a_{n} &=& a_0-a_1=\sum_{q=u,d,s}A^n_q\Delta_q,
\end{eqnarray}
as the results of Eq.(9).

Generally, the spin-dependent  cross section can be decomposed into
the isoscalar $S_{00}$, isovector $S_{11}$ and interference term
$S_{01}$ as
\begin{equation}
S_{SD}^A=a_0^2S_{00}(q)+a_1^2 S_{11}(q)+a_0a_1S_{01}(q),
\end{equation}
where $a_0,\;a_1$ are just the coefficients of isospin 0 and 1 components of the effective current.

In fact, the amplitude for the spin-dependent collision between DM
particle and nucleus is similar to that of nuclear $\beta$ decay
which was thoroughly studied by the nuclear physicists long time ago
\cite{walecka}. The calculations on the SD cross  sections have been
discussed by many authors \cite{many}. Here we just outline the work
given in literature which will be useful for our analysis of the
data taken by the CDEX experiments.

Now let us forward to the stage of DM-nucleus. For zero-momentum
transfer, $$\bar N\gamma_{\mu}\gamma_5
N=u^{\dagger}_N{\mbox{\boldmath $\sigma$}} u_N,$$ where $u_N$ is the
simple two-component fermion spinor of nucleon. When the momentum
transfer is not zero (${\bf q}\neq 0$), the isovecor part of the
axial current induces a pseudoscalar term\cite{engel} as shown in
Eq.(\ref{induce}). By the reduction formula of the QFT, the axial
current should couple to a pseudoscalar meson, and the lightest
pseudoscalar meson is the pion, so that the virtual pion would
compose the most important contribution to the new term. Thus we
have relations $b_0=0$ and $b_1={a_1 m_N \over{\bf q}^2+m_{\pi}^2}$
by assuming PCAC.
Then under the non-relativistic  approximation the matrix element
(\ref{induce})  becomes
\begin{equation}
<p,s|{\cal J}^5_{\mu}(x)|p',s'>= u^{\dagger}_N(p,s)[{1\over 2}(a_0+a_1\tau_3)
{\mbox{\boldmath $\sigma$}}-{1\over 2}{a_1{\mbox{\boldmath $\sigma$}}
\cdot {\bf q}\tau_3\over {\bf q}^2+m_{\pi}^2}{\bf q}]u_N(p',s') e^{i {\bf q}\cdot {\bf x}-i\omega t}
\end{equation}
as $q_0$ being zero for the elastic scattering and for very low
energy inelastic scattering we still can approximate $e^{-i\omega
t}\approx 1$.

The total cross section commonly is  evaluated in terms of the
multipole operators method \cite{donnelly}. Under the
non-relativistic approximation, the scattering amplitude can be
written as
\begin{equation}
{\cal M}={\bf l}\cdot \int d^3x <JM|{\bf {\cal  J}}|JM'>e^{i{\bf q}\cdot {\bf x}},
\end{equation}
thus the differential cross section reduces into
\begin{equation}
{d\sigma\over dq^2}={G\over (2J+1)v^2}S(q),
\end{equation}
where
\begin{equation}
S(q)=\sum_{L\; odd}(|<J||{\cal T}_L^{el}(q)||J>|^2+|<J||{\cal L}_L(q)||J>|^2),
\end{equation}
and $G$ is a constant depending on the model adopted for the
calculation, ${\cal J}^{el}(q)$ and ${\cal L}(q)$  are the
transverse and longitudinal  electric projections of the axial
current. The explicit expressions of ${\cal T}^{el}(q)$ and ${\cal
L}_L(q)$ are defined and calculated in
Refs.\cite{engel,engel:1991wq,ressell:1993qm,ressell:1997kx}, for
saving space, we do not present them here and the readers who are
interested in the details are recommended to refer the original
works.

It is also noted that since the matrix  elements depend on the
expectation values of ${\mbox{\boldmath $\sigma$}}$, the
contributions from the nucleons at lower energy states, in the
language of the shell model, i.e. the nucleons at the inner shells,
should cancel each other. In other words, the nucleons at lower
states would make null contributions to the scattering matrix
elements. Only a few nucleons residing on the very outer shells, the
so-called valence nucleons make substantial contributions to the
DM-nucleus scattering. That is why the SD cross section is much more
difficult to be measured than the SI one.

It is interesting to look deeper on the SD transitions because there
may occur inelastic scattering processes. As aforementioned, the
gaps between the energy levels pertaining to the different principal
quantum numbers are large compared to the available kinetic energy
of the dark matter particle, so that the nucleon which is colliding
with the DM particle cannot transit to an energy shell with higher
principal quantum number. However, due to existence of the ${\bf
L}\cdot{\bf S}$ coupling, the energy level ($l\neq 0$) which was
degenerate would be split into two levels, and the gap between the
two levels is small and completely comparable with the kinetic
energy of DM particle, so that the nucleon may transit into a higher
energy level after the collision and the scattering is inelastic.

Moreover, via the loop effect, the axial  current Lagrangian can
induce an effective scalar coupling which is SI-type and can be
enhanced by the factor $A^2$. Thus the effective coupling is
loop-suppressed, but the total cross section is enhanced by $A^2$,
so its net effect may be comparable with that of the tree
contribution of the axial currents. The situation becomes more
complex and needs to be carefully investigated when the data are
analyzed.

Anyhow, extracting useful information  from the data is by no means
an easy job. Not only careful analysis on the background, but also a
serious theoretical study must be carried out.

As introduced above, the accuracy for detecting the SI cross section
has already reached $10^{-44}$ cm$^2$, it is well known that if the
cross section is smaller than $10^{-46}$ cm$^2$, the contribution of
the atmospheric neutrinos cannot be eliminated. Thus if to
$10^{-46}$ cm$^2$ (the accuracy might be reached in  a few years),
the DM flux is still not observed there could be some possibilities,
one is that the DM particles do not interact with the SM particles
via weak interaction, and another possibility is that  the DM
particles are no WIMPs, but something else, for example heavy
sterile neutrino etc. If it is the first, DM only participates in
gravitational interaction, it would be a disaster for us because on
the earth, the present available apparatus has no chance to measure
gravitational effects, and in the future it is an unanswered
inquiry. If it is the second, we need to do more theoretical study
to explore possible channels to check the postulates.

For the SI cross section, because the Hamiltonian caused by the
effective interaction does not contain a spin-operator, the spin of
the nucleon cannot flip during the collision, thus the scattering is
fully elastic. By contraries, for the SD cross section, the
Hamiltonian  contains a spin operator which may induce a spin flip,
therefore the spin projection of the final state of the nucleon may
be different from the initial one, namely a transition from lower
energy level to higher one occurs during the collision. Obviously
the nucleus is excited after the inelastic collision, then it
definitely will return back to the ground state by radiating a
photon. The photon should have a characteristic spectrum, which can
be "seen" in a detector. The signal might be weak and detection is
rather difficult as expected. This scheme would be a subject of
further investigation. However, it is not applicable to our Jin-Ping
project because the light signals are not detected at our germanium
detector at all.

Indeed, we prey that the DM particle  indeed interacts with SM
quarks, then nucleons via weak interaction, so that we can find its
trace through the direct search on the underground  detectors,
otherwise, one would be unable to identify the mysterious matter
even though we know for sure of its existence.
\section{Review of experiment}

\subsection{Overview}
It is very important for scientists to detect dark matter particles
with different experimental methods in order to understand the
essence of dark matter. Usually experimental detection of dark
matter can be divided into two types: direct detection and indirect
detection. For indirect detection of dark matter, the  SM particles,
such as gammas, neutrinos, and etc., which are generated from
annihilation of two dark matter particles or decays of DM particles,
can be detected. The possible site for dark matter particles to
annihilate each other should be the vicinity near the center of a
star such as the Sun or planets including our Earth, where the
gravitational force traps the dark matter particles whose density
rises to a relatively high level, so that the possibility of dark
matter particle annihilation there would be higher than at other
places. The large-scale high-energy  accelerator may also detect
indirectly the dark matter particles which are generated by
collisions of energetic particle beams. But since the produced dark
matter particles are stable neutral bosons (might be fermions also),
so they only manifest as missing energy, detecting them is a
challenge to our detection technology. Instead, direct detection of
dark matter particles mainly focuses on the measurement of the
deposited energy of a recoiled nucleus scattered off by the coming
dark matter particle, mainly the Weakly Interacting Massive Particle
(WIMP). The essence of direct detection of WIMP is to single out the
possible events induced by the coming dark matter particles from a
large background produced outside and inside the detector system. So
the ultra-low background level of the detecting system for direct
detection of WIMP is a crucial requirement Fig.\ref{fig1}.
\begin{figure}[h]
\vspace*{0.2cm}\caption{~~The principle of WIMP direct detection.\\~} \label{fig1}
\hspace*{-0.5cm}\begin{minipage}[h]{\textwidth}
\includegraphics[scale=1.1]{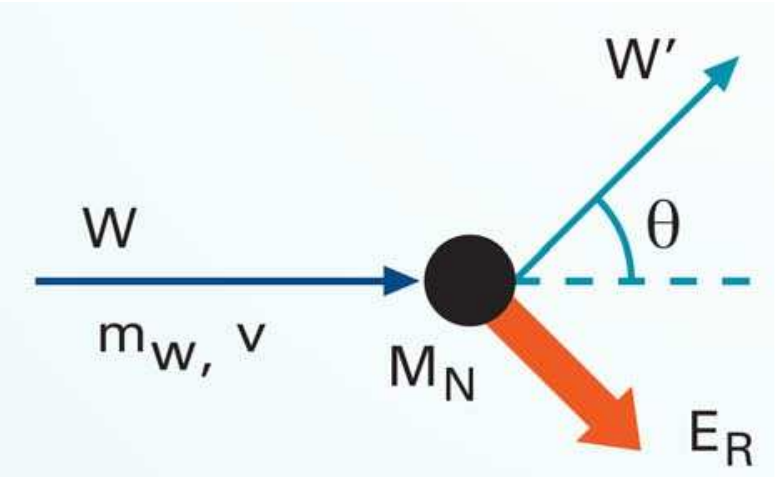}
\end{minipage}
\end{figure}

For different detectors, the deposited energy of the recoiled
nucleus scattered off by the incident WIMP in the fiducial volume of
the detector usually can be detected with three different processes:
ionization, scintillation and heating. So the three main  types of
detectors are ionization detector, scintillation detector and
heating detector which measure respectively ionized electrons,
scintillation light and temperature variations. Some detectors are
designed to extract information from a combination of two such
processes in order to improve the ability of discriminating  the
"real" WIMP signal from the background events.

More than tens of groups in the world now are carrying out
experiments to directly detect WIMPs and the results have been
achieved to the level less than $10^{-7}$pb for WIMP mass of around
50 GeV.

For the scintillation method, crystals such as NaI(Tl) and CsI(Tl)
are used as the target detectors. Scintillation detector can provide
a pulse shape  to discriminate the WIMP events from background
events and it is easy to produce a larger prototype detector which
can be a ton-mass scale detector. Ultra-pure scintillation detectors
have been studied and run for a long time to detect WIMP. The DAMA
Collaboration has built its NaI scintillation crystal array detector
with the mass scale of hundreds of kilograms \cite{bernabei}. The
KIMS group has chosen almost the same technology with DAMA except
the scintillation crystal is CsI(Tl) \cite{kims2007}. The relative
low photoelectron yields restrict the application of the pulse shape
discrimination method at low energy region.  Both the DAMA and KIMS
experiments  turn to detect the annual modulation of the event rate
induced by WIMPs(Fig.\ref{fig2}). The DAMA Collb. has achieved an
average event rate of 1cpkkd above 2 keV and given a clear annual
modulation result with 1.17 ton$\times$yr data set and 13 annual
cycles. Based on these results, the DAMA collaboration claimed that
they had found evidence of dark matter.
\begin{figure}[h]
\vspace*{0.2cm}\caption{~~The relative velocity of the earth to the sun and the velocity of the sun moving in the WIMP "sea".\\~} \label{fig2}
\hspace*{-0.5cm}\begin{minipage}[h]{\textwidth}
\includegraphics[scale=1.1]{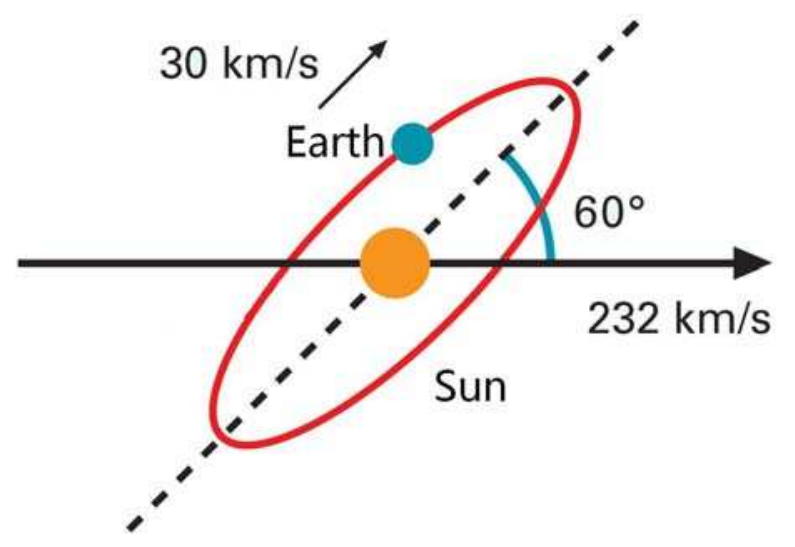}
\end{minipage}
\end{figure}

For the ionization method, high purity germanium detector is the
best choice till now due to its ultra-low self-radioactivity and
feasibility to develop large-scale detector technically. Germanium
detector could provide low energy threshold and very good energy
resolution, but its shortcoming is a not-good ability to
discriminate the recoiled events induced by incident WIMPs from the
background events of gamma and electron. This disadvantage limits Ge
detectors or other semiconductor detectors to reach high detection
accuracy for a relatively large background.

For the heating method, the detectors could measure a tiny vibration
of temperature and heat deposition inside of the detector materials
when the detectors run in the ultra-low temperature circumstance
with only tens of millikelvin. The heat deposition can induce a
change of the current signal in the equipped electronics and then
will be recorded.

For dark matter search experiments, the primary and key task is to
extract out possible signals  from the recorded events which include
a large number of background debris and determine if they are
induced by the incident dark matter particles. So several
experimental groups choose detectors which could simultaneously
measure two kinds of signals induced by one interaction and this
strategy could make the detector to possess a very strong ability to
discriminate background. Now two collaborations, CDMS and XENON,
have published the most stringent and sensitive results of a
detection for tens of GeV-mass  WIMP. The CDMS group has developed a
new type of Ge and Si detector which collects both the ionization
and heat signals. A kind of superconducting tungsten transition-edge
sensors (TESs) has been used to read out the heat deposition, at the
same time the ionization signal is also recorded. So two kinds of
signals including ionization and heat are read out when one incident
particle hits the target in the fiducial volume of the detector. The
ration of the ionization energy and heat energy could be different
for the recoils of the nucleus which is  scattered off by WIMP and
the background events caused by incident gammas and/or electrons.
This scheme provides a very strong event discrimination for the CDMS
experiment. The CDMS group has run its Ge and Si detecting system
for several years and published its new observation result in 2010
and 2011 (Fig.\ref{Fig3}).
\begin{figure}[h]
\vspace*{0.2cm}\caption{~~The physical results from CDMS, XENON and other group for WIMP search.\\~} \label{fig3}
\hspace*{-0.5cm}\begin{minipage}[h]{\textwidth}
\includegraphics[scale=1.1]{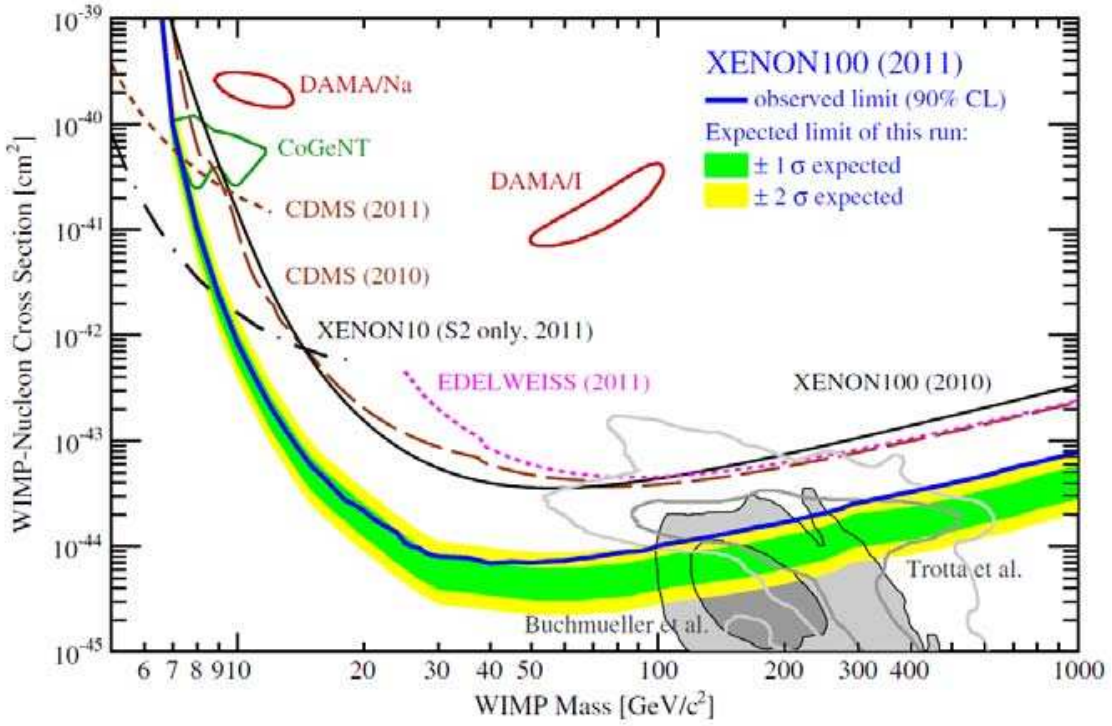}
\end{minipage}
\end{figure}

The XENON group has also developed their liquid Xenon TPC( Time
Project Chamber) detector which collects both the ionized electrons
and scintillation light when a particle interacts with the xenon
target. Now the XENON100 experiment has built its liquid xenon
detector of 100 kg fiducial mass and published new results in 2011
\cite{sciencexpress2010}.
\subsection{Ultra-low energy threshold experiment}

In recent years, an ultra-low energy threshold of about 400 eV has
been achieved with the germanium detector based on the point-contact
technology, the point-contact germanium detector PCGe  is used for
scanning WIMP of mass as low as 10 GeV. The scientists from Tsinghua
university of China first started the experimental preparation from
2003 \cite{yueq2004} and ran the first 5g-mass planar Ge detector
for a test. The TEXONO collaboration runs a 20g ULE-HPGe detector
with the shielding system on the ground near a nuclear power plant
in Taiwan and published its physical results in 2009 shown in
Fig.\ref{fig4} \cite{texono2009}. The CoGeNT Collaboration has also
started a similar experiment for the WIMP search with 475g PCGe
detector at a ground laboratory from 2006 and published its physical
results in 2008 and 2011, respectively \cite{cogent2011}. The CoGeNT
experiment has explored a low WIMP mass region and its results are
shown in Fig.\ref{fig5}.
\begin{figure}[h]
\vspace*{0.2cm}\caption{~~The physical results of WIMP detection from TEXONO group.\\~} \label{fig4}
\hspace*{-0.5cm}\begin{minipage}[h]{\textwidth}
\includegraphics[scale=1.1]{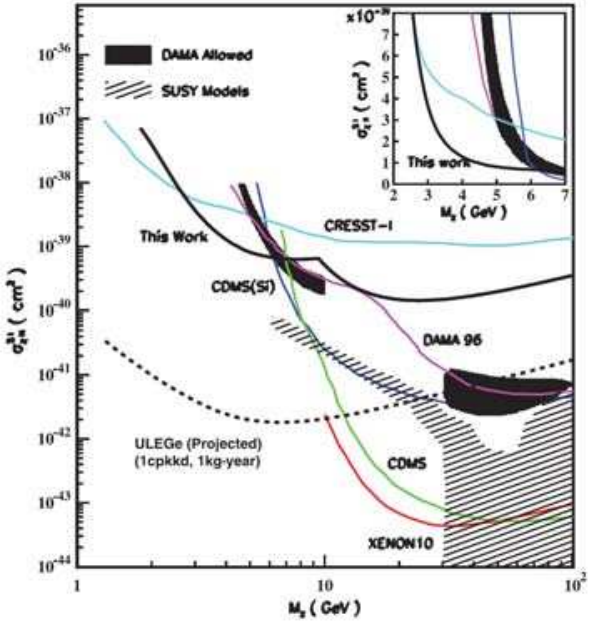}
\end{minipage}
\end{figure}
\begin{figure}[h]
\vspace*{0.2cm}\caption{~~The physical results of WIMP detection from CoGeNT group.\\~} \label{fig5}
\hspace*{-0.5cm}\begin{minipage}[h]{\textwidth}
\includegraphics[scale=0.6]{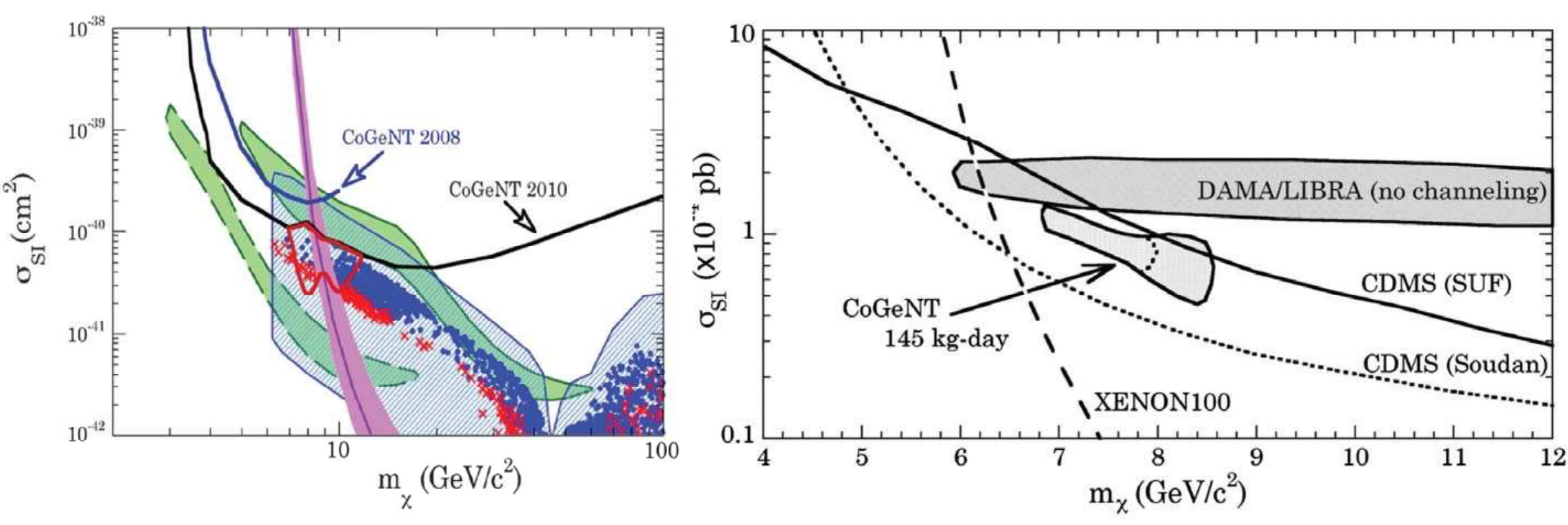}
\end{minipage}
\end{figure}

Due to the updated results from Ge detectors, the low-mass WIMP
detection has become one of the new hot topics for dark matter
experiments. Many groups have tried to develop lower threshold
detectors to scan the low mass WIMP region below 10 GeV. The
Majorana\cite{majorana} and GERDA\cite{gerda} groups try to reduce
their energy threshold to cover the energy range for both double
beta decay and dark matter search.

In 2009, the CDEX Collaboration is aiming to search for low mass
WIMPs with a ton-scale point-contact germanium detector located at
the Jin-Ping underground Laboratory (CJPL) in China. Now the first
stage of the CDEX project-CDEX-1, where one 1kg-mass PCGe detector
is installed has already successfully run and begun to take data.

\section{The CJPL}

Experiments such as detection on dark matter, double beta decay
experiment and neutrino experiment in the particle physics domain
require Ultra-low background laboratories in order to identify the
rare events. Those deleterious background events come mainly from
the radioactive isotope of environmental materials, high energy
cosmic-ray muons which originate from the interaction of high energy
protons in cosmic rays with atmospheric compounds at outer space, as
well as the internal backgrounds from active isotopes in detector
materials and the noise of the detector electronics. The ambient
radioactive background from the radioactive nuclei could be shielded
with efficient shielding system which includes possible passive and
active shielding parts. However the high energy muons, which are the
main hard contents of cosmic-ray, can pass through the environment
and interact with the shielding system materials, the structure
materials and the detector itself. Though the instantaneous events
coming from direct interaction of muons with those materials on
their paths can be vetoed by an active muon veto system, the delayed
neutron and accompanying radioactive nuclei induced by incident
cosmic-ray muons can contribute to backgrounds in the detector. This
channel is the most difficult one to be shielded and extracted out
from the spectrum of the detection. So it is very necessary that the
detection of dark matter should be performed at underground
laboratories where the muon particles are efficiently stopped and
absorbed by the overburden rock.

There are many underground laboratories established or under
construction in the world including the LNGS in Italy, Kamioka in
Japan, Subdury in Canada, Modane in France, Soudan in USA, and so on
\cite{taup2009}. In 2010, the first deep underground laboratory was
built with excellent working and living conditions in China. This
deep underground laboratory is named The China Jin-Ping Underground
Laboratory (CJPL).

\subsection{The CJPL environment}

The Yalong River is  more than 1500-km long in Sichuan province of
China. A part of almost 150 km of the river bends and encompasses
the huge Jin-Ping Mountain to make a narrow arc. The turning is very
sharp at the turning point, namely in the mathematical terminology
the curvature radius is small at the spot, thus the west and east
parts of the river are not much apart, but separated by the mountain
and the height difference of their water surfaces is quite large. If
a tunnel is drilled from the east to the west along the intercept of
the arc, the water drop height is tremendously large, so that this
can serve as an ideal hydrodynamic resource. Two hydropower plants
at each side of the Jin-Ping Mountain on the Yalong River are being
built now by Yalong River Basin Hydropower Development Company. Totally seven
parallel tunnels are drilled including one drainage tunnel, two
transport tunnels and four headrace tunnels\cite{cjplscience}. In
2008, these two transport tunnels were completed and have been in
use for the hydropower plant project, and the map is shown in
Fig.\ref{fig6}. The length of these two transport tunnels is 17.5 km
and the cross-section is about 6m$\times$6m. The CJPL is located in
the central portion of one of the transport tunnels and the rock
overburden is about 2400 m thick. Fig.\ref{fig7} shows the detailed
location of the CJPL in the transport tunnel and the transect
profile of the transport tunnel.
\begin{figure}[h]
\vspace*{0.2cm}\caption{~~The site of CJPL.\\~} \label{fig6}
\hspace*{-0.5cm}\begin{minipage}[h]{\textwidth}
\includegraphics[scale=1.1]{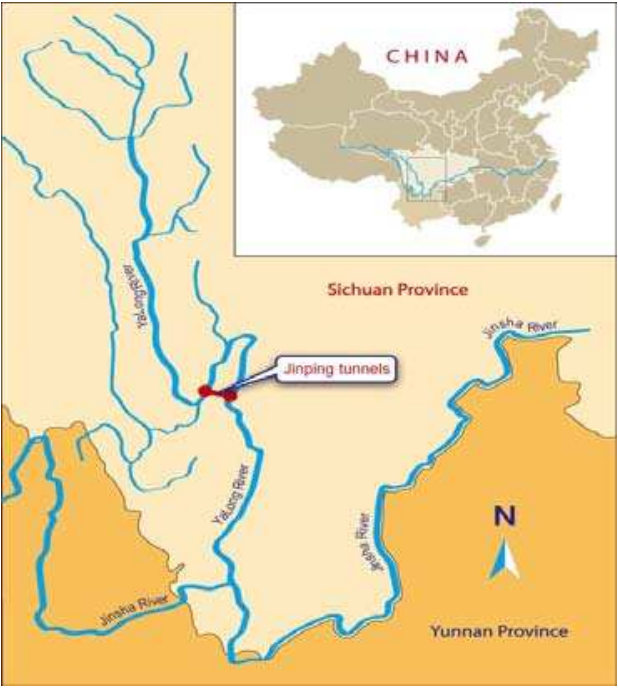}
\end{minipage}
\end{figure}
\begin{figure}[h]
\vspace*{0.2cm}\caption{~~The cross-section of the Jin-Ping Mountain along the transportation tunnel.
The site of CJPL is in the middle part of the Jin-Ping tunnel and can access by drive.\\~} \label{fig7}
\hspace*{-0.5cm}\begin{minipage}[h]{\textwidth}
\includegraphics[scale=0.8]{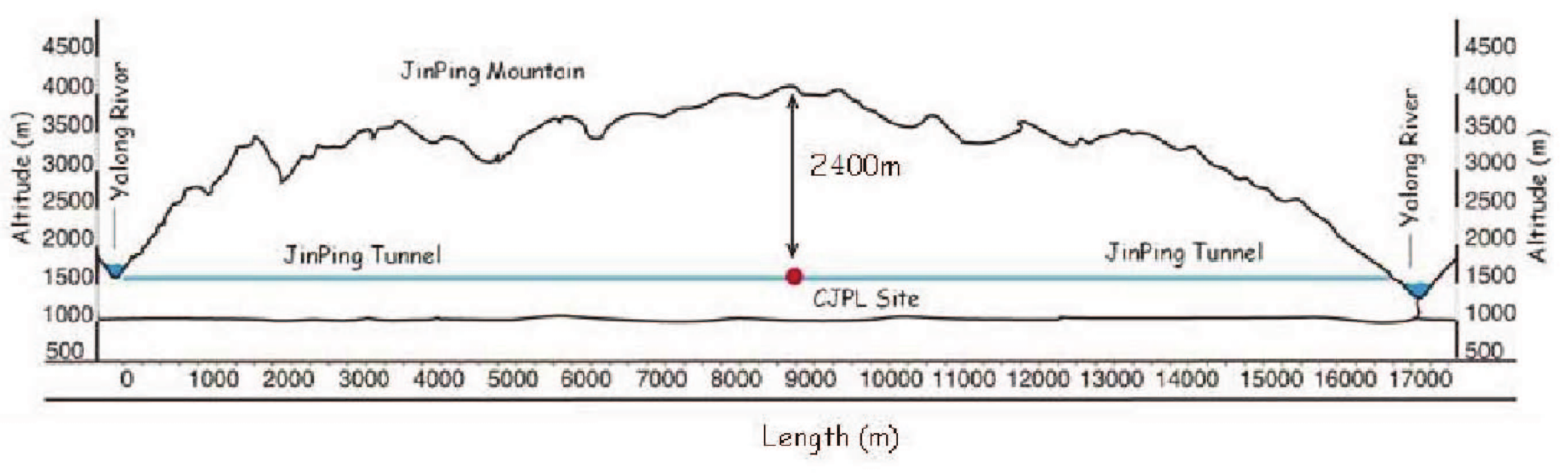}
\end{minipage}
\end{figure}

Tsinghua University, collaborating with Yalong River Basin (originally called Ertan) Hydropower Development
Company who owns the Jin-Ping tunnel, had made a plan to build an
underground laboratory which was large enough to host a relatively
large-scale low-background experiment. The project is under way
right now. As the first step, a small CJPL hall has been constructed
for dark matter experiment and an ultra-low background material
screening facility has also been installed. The CJPL internal space
includes three parts: the entrance tunnel of 20 m in length, 30
m-long connection tunnel and the main hall with dimension of
7.5m(H)$\times$6.5m(W)$\times$40m(L) and the total available volume
is about $\mathrm{4000m^3}$. The wall of CJPL is covered by a layer
of air-proof resin to separate it from the rock of the tunnel. There
are a ground laboratory building, office and dormitory for the
researchers near the entrance of the tunnel. Apartments,
restaurants, hotel and sport facilities are also available nearby
the ground laboratory.

In order to provide a  good working condition with fresh air for
researchers and further decrease  the radon concentration in the air
of the internal space, a 10-km long air ventilation pipe has been
built to pump the fresh air from outside the transport tunnel into
the CJPL space. This ventilation system can provide up to
2000$\mathrm{m^3/h}$ fresh air and keep the air clean inside the
CJPL. According to the need of dark matter experiment, the radon
trapping system will be installed in the CJPL serving as an improved
radon gas filter. The CJPL is equipped with 3G wireless network and
fiber access to broad-band internet.

\subsection{CJPL facilities}

Low background germanium spectrometer serves as the standard
facility for material screening and selection for either detection
of dark matter or neutrinoless double beta decay experiments
\cite{heusser}. A low background germanium spectrometer, called
GeTHU facility, with a dedicated low background shield has been
designed and set up lately at CJPL for material selection of dark
matter experiment CDEX and other rare-event experiments. Now the
facility is operating for background measurement. Moreover, another
two counting facilities are being designed to improve minimum
detecting sensitivity.

The detector is a high-purity, N-type germanium detector (HPGe) with
a relative efficiency of 40$\%$ and was constructed by CANBERRA in
France \cite{canberra}. The germanium crystal has a diameter of 59.9
mm and a thickness of 59.8 mm. The cryostat is made of ultra low
background aluminum with a U shape to avoid direct line-of-sight to
outside (see Fig.\ref{fig8}(a)). The preamplifier is placed outside
the shield, since it causes more radioactive contaminations.

The shielding structure has been designed to guarantee a large
sample space, low background and easy operation. The sample chamber
is surrounded by 5 cm (15 cm for the base plate) of oxygen-free,
highly pure copper made by the Chinalco Luoyang Copper Co., Ltd
\cite{luoyang}. Three layers of ordinary lead, each 5 cm thick,
surround the copper(see Fig.\ref{Fig8}(a) and (b)). The $\mathrm{^{210}Pb}$
activity of lead is about 100 Bq$/$kg. All lead bricks were
carefully cleaned by ethanol before installing. Outside the lead are
10 cm borated polyethylene plates to prevent penetration of ambient
neutrons. The upper copper plate, closing the sample chamber,
carrying the upper lead bricks and borated polyethylene plates, are
placed on sliding rails in order to open or close the shield easily.
The whole system is flushed by boiling nitrogen from the cooling
dewar. The Monte Carlo simulation shows the internal background
count rate of GeTHU within 40 keV to 2700 keV is only 0.0007 cps or
so.(see Fig.\ref{fig8}(c))
\begin{figure}[h]
\vspace*{0.2cm}\caption{~~The ultralow background HPGe gamma spectrometry at CJPL.\\~} \label{fig8}
\hspace*{-0.5cm}\begin{minipage}[h]{\textwidth}
\includegraphics[scale=1.1]{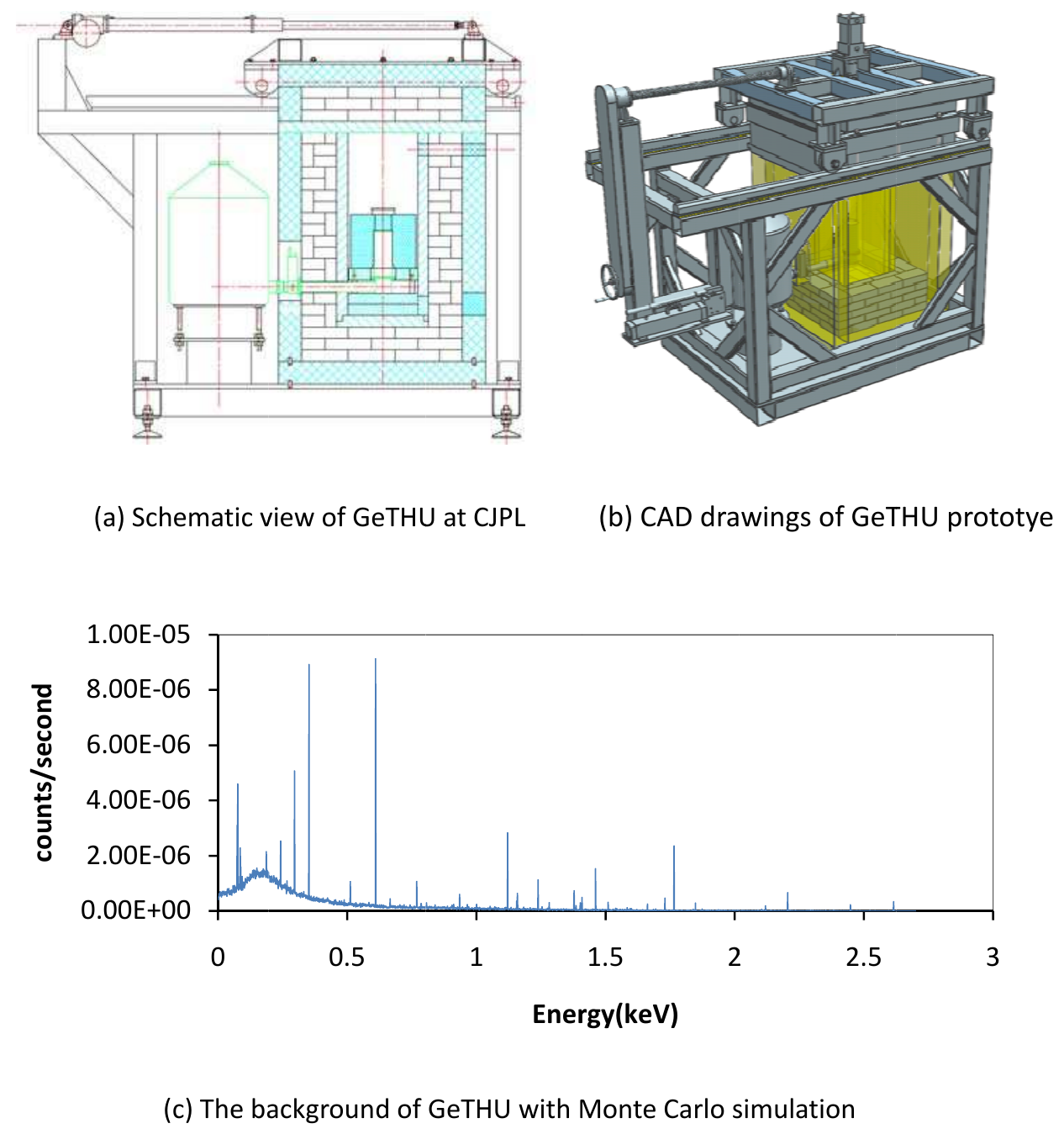}
\end{minipage}
\end{figure}

\subsection{CJPL key performance(specification) $+$ simulation}

\subsubsection{The Radioactivity of surrounding environment at CJPL}

Original rock samples at different positions in the cave were
collected before construction and the concrete samples were
collected during the construction. All samples were measured and
analyzed by low-background HPGe gamma spectrometer. The measurement
results are following: the radioactivity concentrations of
$\mathrm{^{226}Ra,~^{232}Th}$ and $\mathrm{^{40}K}$ of rock samples
are $\mathrm{1.8\pm 0.2Bq/kg, <0.27Bq/kg}$ and $\mathrm{<1.1
Bq/kg}$; and that of concrete samples are $\mathrm{1629.3\pm 171.3
Bq/kg, 6.5\pm 0.9Bq/kg}$ and $\mathrm{19.9¡À3.4 Bq/kg}$.

Beside the sample measurement, a portable gamma spectroscoper
manufactured by ORTEC$^{\mathrm{TM}}$ is used to characterize
dispersed radioactive nuclides in the environment at CJPL areas.  A
portable HPGe spectrometer is used to measure the gamma flux at
CJPL. The detector consists of a HPGe crystal with a mass of 709 g,
the signals from which will be stored and analyzed by the
ORTEC$^{\mathrm{TM}}$ DigiDART multichannel analyzer(MCA). The
spectrometer is set up to measure gamma ray with energies up to 3
MeV£¬and the energy resolution(FWHM) is 1.6 keV at 1.33MeV.
Fig.\ref{fig9} shows the in-situ gamma spectrum in different places.
\begin{figure}[h]
\vspace*{0.2cm}\caption{~~In-situ Gamma spectra in 3 places: CJPL Hall-green line,
CJPL PE shielding house-purple line, Ground laboratory outside Jinpin Tunnel -blue line.\\~} \label{fig9}
\hspace*{-0.5cm}\begin{minipage}[h]{\textwidth}
\includegraphics[scale=1.1]{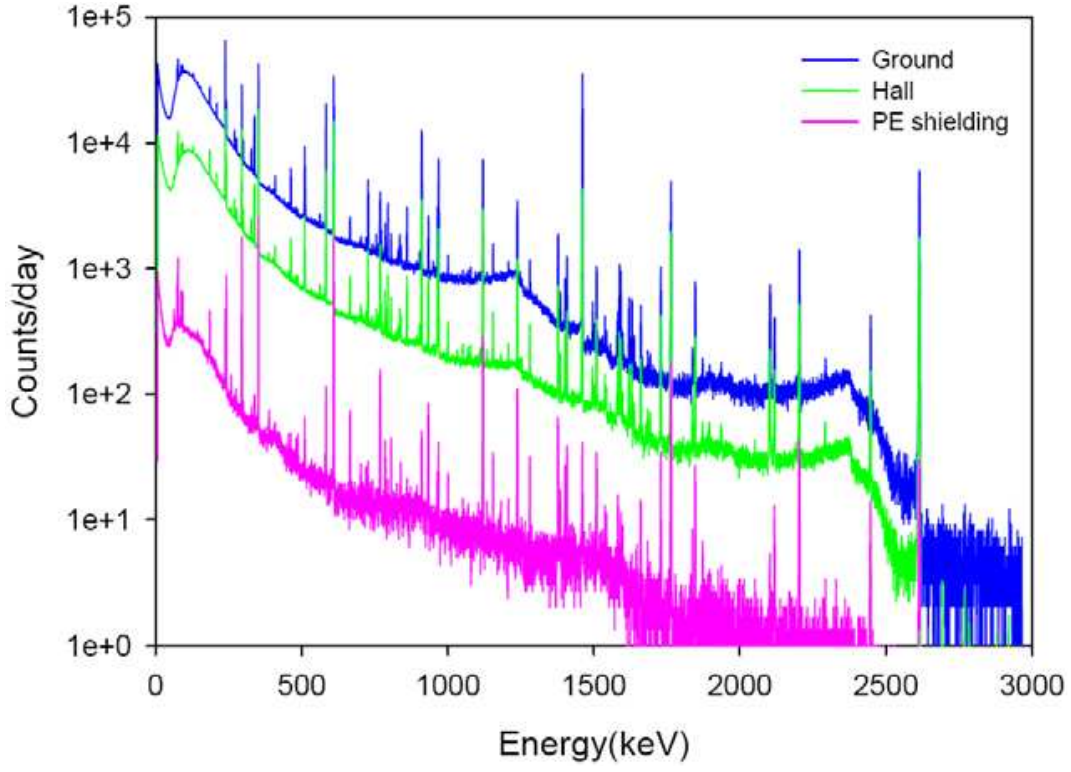}
\end{minipage}
\end{figure}

\subsubsection{Cosmic Ray (Muon)}

The main components of cosmic ray which can pass through rock
stratum of the mountain are muon and neutrino. The neutrino
component can be ignored because of its small interaction cross
section with the detector materials. However, it's very important to
know the exact flux of cosmic-ray muons for estimating the
background event rate caused by cosmic ray directly or indirectly.

In order to obtain the exact value of the muon flux in CJPL, 6
plastic scintillation detectors are employed. The volume of plastic
scintillators  is 1m$\times$0.5 m$\times$0.05 m. 6 detectors are
divided into groups A and B. Each group has three plastic
scintillation detectors. The three plastic scintillation detectors
of each group are put in erect on a shelf. The up-down distance from
one detector to the neighbor one is 0.20 m. The gap between the two
groups is about 0.20 m. Data are taken by a DAQ system and the
LabView is shown in Fig.\ref{fig10}.
\begin{figure}[h]
\vspace*{0.2cm}\caption{~~Schematic diagram of cosmic ray muon detect system.\\~} \label{fig10}
\hspace*{-0.5cm}\begin{minipage}[h]{\textwidth}
\includegraphics[scale=0.6]{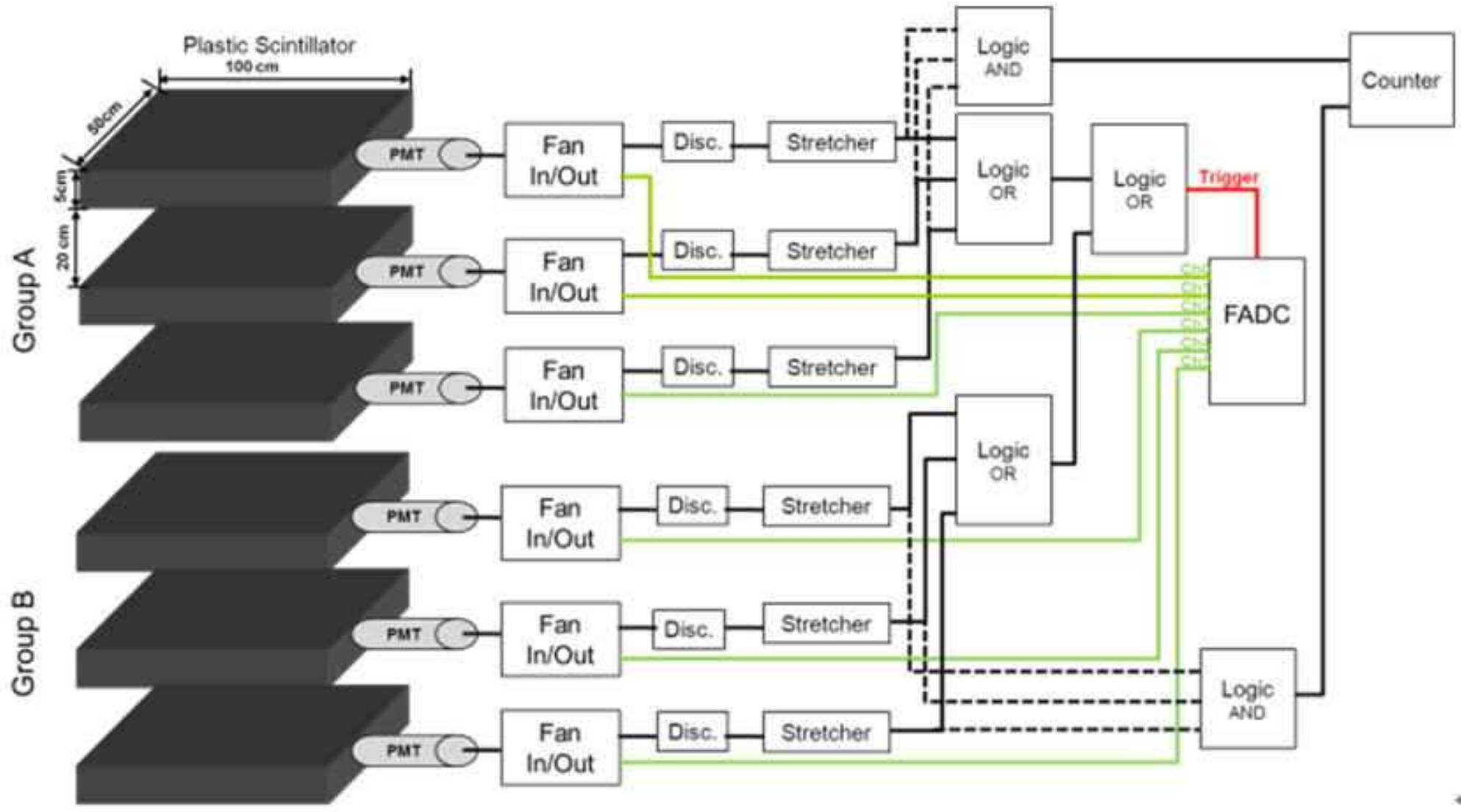}
\end{minipage}
\end{figure}

The performance of the whole detection system was investigated on
the ground nearby entrance of east end of the tunnel before
measurements began inside CJPL. The triple coincidence event of the
3 plastic scintillation detectors in one group is caused by cosmic
ray muons. The signal of muon event is much larger than the noise,
so the experimental environment is very clean.

The preliminary result is 0.16  per square meter per day, i.e. about
1 of $10^8$ muons on the ground.


\subsubsection{Radon Monitoring in CJPL}

A decay product of $\mathrm{^{238}U}$, the noble gas
$\mathrm{^{222}Rn}$ as well as its decay daughters, also
'observably' contribute to the natural background radioactivity in
underground laboratories as it is emanated from the rock and can
rather easily enter the detector. Variations of the air radon
concentration both in the hall and shielding house are continuously
measured using two Alphaguard radon monitor manufactured by SAPHYMO
GmbH (see Fig.\ref{fig11}). During long period monitoring, the
average air radon concentration is $\mathrm{\sim 100 Bq/m^3}$
without ventilation and ~50 $Bq/m^3$ with ventilation.
\begin{figure}[h]
\vspace*{0.2cm}\caption{~~Radon Monitor in CJPL experiment Hall.\\~} \label{fig11}
\hspace*{-0.5cm}\begin{minipage}[h]{\textwidth}
\includegraphics[scale=1.1]{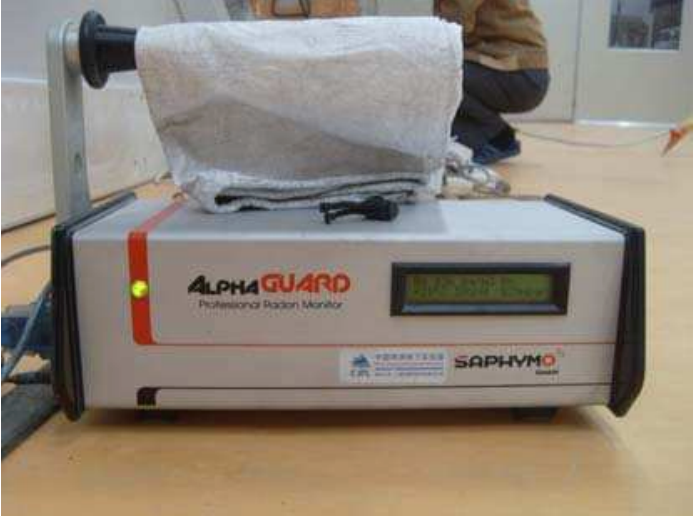}
\end{minipage}
\end{figure}

\section{The CDEX experiment}

\subsection{Introduction to CDEX }

The main physics goal of the CDEX project is to search for WIMPs in
a mass range of 10 GeV with sensitivity better than
$\mathrm{10^{-44}cm^2}$. Because of much more advantages, such as
low radioactivity, high energy resolution, high matter density and
stability at work, the CDEX adopts the High purity point contact
Germanium PCGe as the target and detector. The recoil energy of
Germanium nucleus for low mass WIMPs is only several keV (
Fig.\ref{fig12}). Considering the quench factor the threshold of
detection should be 100-300 eV for the WIMPs within the range less
than 10 GeV. The mass target or detector should be as large as
possible because of low event rates. The detector mass of the first
phase CEDX-1 is 1 Kg and that of the second phase CDEX-10 is 10 kg.
The final goal of the CDEX project is to set up to a ton-scale mass
Ge detector CDEX-1t in the CJPL.¡¡
\begin{figure}[h]
\vspace*{0.2cm}\caption{~~The recoil energy of PCGe for deferent WIMP mass.\\~} \label{fig12}
\hspace*{-0.5cm}\begin{minipage}[h]{\textwidth}
\includegraphics[scale=1.1]{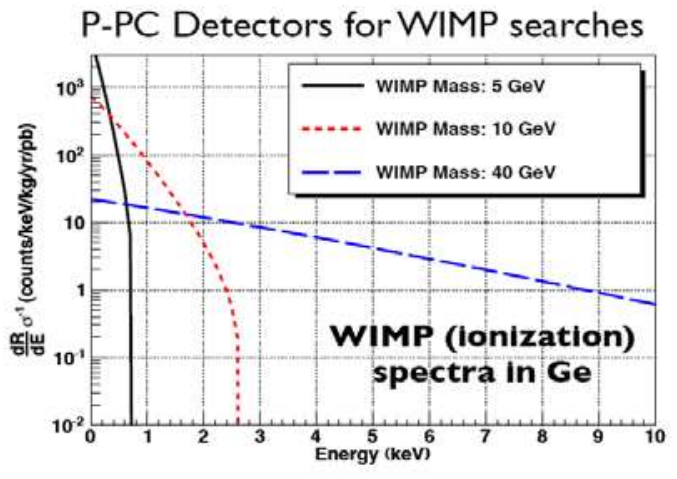}
\end{minipage}
\end{figure}

Obviously, the CDEX is a kind of ultra-low-background experiment.
The biggest challenge is to reduce the background event rate to an
acceptable level. The expectation of background count level is less
than $\mathrm{0.1cpd/keV/kg}$. To ensure its successful operation,
the CDEX must:
\begin{itemize}
\item hold the detector in the CJPL deep underground laboratory;
\item  establish an efficient shielding system to reduce the background further;
\item  design a large mass and low threshold detector with tiny amount of internal radioactive isotopes.
\end{itemize}

As mentioned above, the CJPL, which has an overburden of
about 2400m rock, provides an ideal place to host the CDEX. In the
following, we are going to briefly introduce the detector and
the shielding system.
\subsubsection{Detector}

The physical aim of the CDEX composes an unprecedented challenge to the
design of the CDEX detector. In summary, the detector should have
the following features:
\begin{itemize}
\item 1 Kg to 1000 kg of detector mass to compensate the extremely low cross-section, namely substantially increase the event rate;
\item 200-400 eV threshold; This is because the recoil energies are low,
and furthermore only a fraction of these (a low energy quenching
factor for Ge recoil of O(20)$\%$) is generally in a detectable
form, such as ionization.
\item very low concentration of internal radioactive isotopes.
\end{itemize}

As a semiconductor detector made of the purest material on the
earth, the HPGe detector was first proposed to detect WIMP by our
group in 2004. By using arrays of commercially available HPGe diodes
(5 g, 1pF capacitance), TEXONO achieved 300 eV  energy threshold.
But further increase of the readout channels and monetary investment
prevents the detector mass to reach an O(kg) scale.

In order to compromise the mass constraint and energy resolution,
small semiconductor detectors usually are the conventional HPGe
coaxial detector with low noise and threshold, instead, the CDEX
proposes to use a point-contact HPGe detector (herein named PCGe) to
directly detect the ionization effect of the  recoiled Ge nuclei.
Fig.\ref{fig13} shows the configurations of coaxial HPGe (left) and
PCGe (right) detectors. By using point contact technology, a PCGe
detector can reach an order of 1 kg of mass with very small
capacitance (an order of 1pF) and promising intrinsic noise
characteristics. Besides the requirements on mass, energy resolution
and threshold, it also has advantages that its intrinsic ability to
distinguish multi-site from single-site particle interaction and
surface events is remarked. The mass of the detector array can reach
a scale up to O(10)kg or even O(1000)kg.
\begin{figure}[h]
\vspace*{0.2cm}\caption{~~The configuration of Coaxial HPGe (left) and Point-contact HPGe(right).\\~} \label{fig13}
\hspace*{-0.5cm}\begin{minipage}[h]{\textwidth}
\includegraphics[scale=1.1]{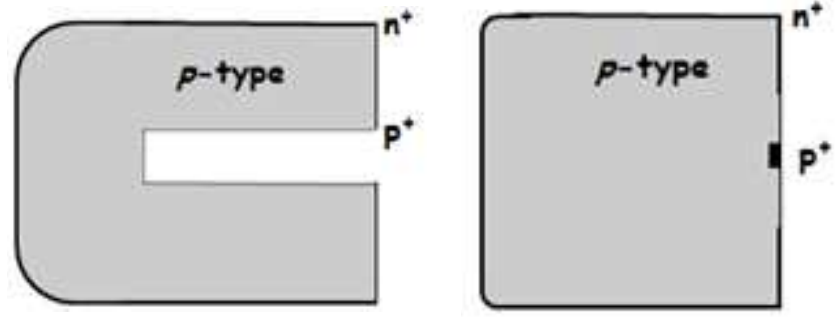}
\end{minipage}
\end{figure}

\subsubsection{Shielding system}

The layout of the CDEX is shown schematically in Fig.\ref{fig14}. It
is located in the CJPL, which has a PE (polyethylene) layer with a
thickness of 1 m built to decelerate and absorb fast neutrons.
Taking into account the commonly accepted WIMP density and elastic
scattering cross sections, the expected event rate is estimated as
$\mathrm{< 0.1kg^{-1}day^{-1}keV^{-1}}$. These extremely rare events
are very difficult to be distinguished from the high backgrounds
which come from cosmic rays and natural radioactivity. Then the
passive and active shielding systems are proposed (shown in
Fig.\ref{fig15}). The outermost layer is a 20 cm lead layer to
reduce the environmental gamma ray radiation. The next layer is a 15
cm steel supporting structure. Then, a 20 cm PE(B)(Boron-doped
polyethylene) layer is used to absorb thermal neutrons. The
innermost layer is 10 cm OFHC (oxygen-free highly-conductive copper)
for absorbing residual rays. The space inside the copper layer is
the room for the HPGe detector and the active shielding system. To
reduce the background of radon gas, the inner space of the shielding
should be refreshed continuously with highly pure nitrogen gas.
\begin{figure}[h]
\vspace*{0.2cm}\caption{~~The layout of CDEX and surrounding rock.\\~} \label{fig14}
\hspace*{-0.5cm}\begin{minipage}[h]{\textwidth}
\includegraphics[scale=1.1]{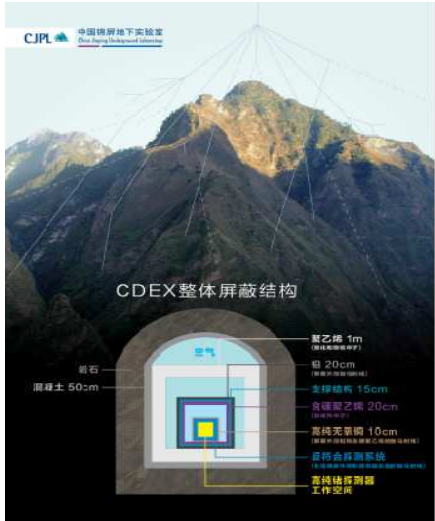}
\end{minipage}
\end{figure}
\begin{figure}[h]
\vspace*{0.2cm}\caption{~~The schematic diagram of the proposed passive shielding system.\\~} \label{fig15}
\hspace*{-0.5cm}\begin{minipage}[h]{\textwidth}
\includegraphics[scale=1.1]{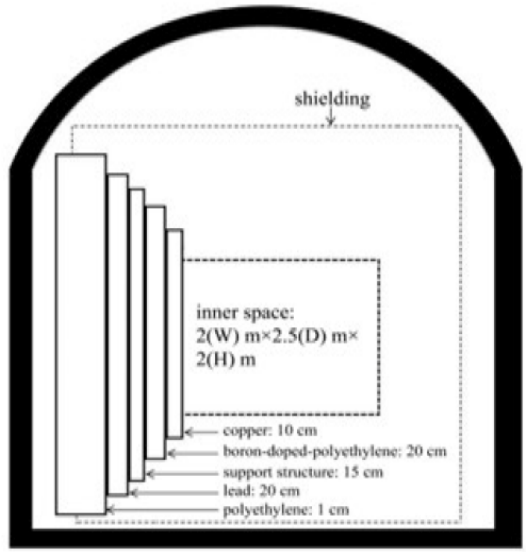}
\end{minipage}
\end{figure}

Based on our Monte-Carlo simulation of environmental radioactivity
and cosmic-ray using Geant4 Monte Carlo codes, the environment of
the CJPL (rocks and concrete): neutrons yielded from the rock and
reaching the innermost region of the passive shielding system is
$\mathrm{3.129\times 10^{-12}cpd}$, the neutrons yielded from the
concrete layer and reaching the innermost region of the passive
shielding is $\mathrm{6.490\times 10^{-10}cpd}$, these two
quantities are much smaller than the criteria  for neutron
background demanded by our expected goal. For the passive shielding
system itself: we mainly consider the neutron background yielded
from the lead and copper shielding, the unit contents of
$\mathrm{^{232}Th}$ and $\mathrm{^{238}U}$ have been simulated, the
result is that the neutrons which are produced by 1ppm of
$\mathrm{^{232}Th}$ in lead and finally reach the innermost region
of the passive neutron shielding are $\mathrm{3.909\times
10^{-5}cpd}$, the neutrons which are produced by 1ppm of
$\mathrm{^{238}U}$ and finally reach the innermost region of the
passive neutron shielding are 5.848cpd. The neutrons which are
produced by 1ppb of $\mathrm{^{232}Th}$ in copper and finally reach
the innermost region of the passive neutron shielding are
$\mathrm{1.480\times 10^{-4}cpd}$, the neutrons which are  produced
by 1ppm of $\mathrm{^{238}U}$ and finally reach the innermost region
of the passive neutron shielding are 2.289cpd. Thus the neutron
background from the environment of the CJPL could be reduced very
efficiently by the shielding of the CDEX. The remaining neutron
background mainly comes from the passive shielding system itself. In
order to realize our expected low background, we have to restrict
the radioactive content of the copper and lead bricks.

Different active shielding methods are proposed for different phases
of the experiments. In the first phase, the CDEX-1, CsI(Tl) or
NaI(Tl), a veto detector surrounding the HPGe detector is proposed.
The Liquid Argon LAr veto is proposed in the second phase CDEX-10.
In this design, the LAr detector serves as both the passive
shielding detector and the low temperature media for the HPGe
detector.

\subsection{CDEX-1}

As the first phase of the CDEX experiment, a detector  with 1 kg
mass scale of HPGe, named CDEX-1, is designed and runs first. It
includes a ready-made 20g ULEGe \cite{stlin2009} (also Ultra-LEGe,
Ultra Low Energy high purity Germanium detector) detector and a
1kg-PPCGe(P-type Point Contact Germanium) detector.

\subsubsection{20g-ULEGe}

The 20g ULEGe (N-type), manufactured by Canberra, France in 2005,
actually consist of 4 duplicate crystal elements. Each 5g element
whose cross section structure shown as Fig.\ref{fig16}, has a
semi-planar configuration with a P+ contact on the outer surface,
and an N+ contact of small diameter. Other surface encircling the N+
contact is passivated to suppress the surface dark current.
Fig.17 depicts the horizontal cross section of the cryostat
and the positional relationship of the 4 crystals. The
cryostat end cap opens a 0.6mm thick carbon composite window so that an
external soft X-ray calibration can be carried out. Near the N+
contact, a low noise FET is installed which reads the signal and
inputs it to a pulsed optical feedback preamplifier.
Each crystal element has two
identical outputs, which are connected to high impedance input of
the downstream modules.
\begin{figure}[h]
\vspace*{0.2cm}\caption{~~20g-ULEGe detector geometry.\\~} \label{fig16}
\hspace*{-0.5cm}\begin{minipage}[h]{\textwidth}
\includegraphics[scale=1.1]{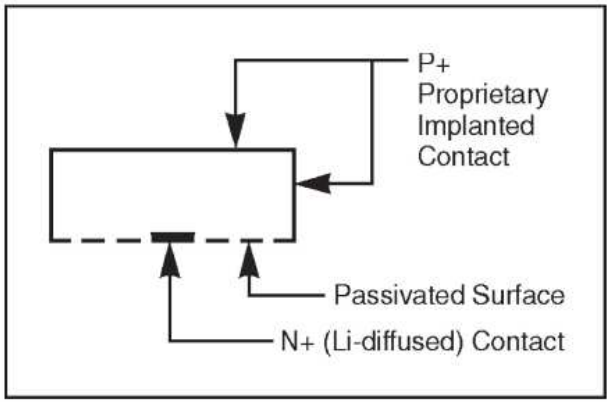}
\end{minipage}
\end{figure}
\begin{figure}[h]
\vspace*{0.2cm}\caption{~~ The horizontal cross section of cryostat and crystal array.\\~} \label{fig17}
\hspace*{-0.5cm}\begin{minipage}[h]{\textwidth}
\includegraphics[scale=0.6]{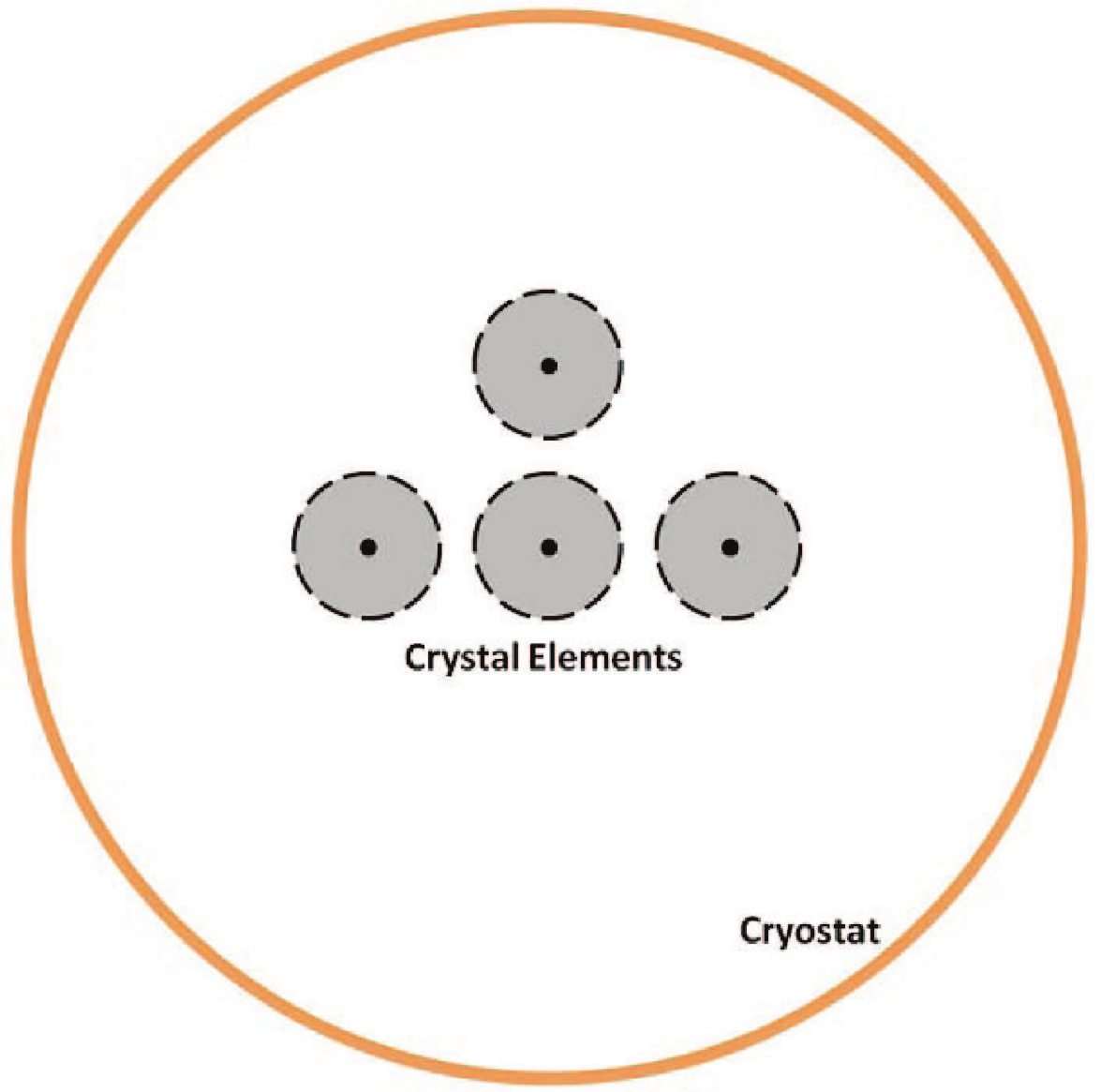}
\end{minipage}
\end{figure}

Fig.\ref{fig18} simply describes the DAQ setup of the 20g-ULEGe
experiment. The output from the preamplifier is directly connected
to the conventional spectroscopy amplifier (Canberra 2026), which
has high input impedance. The signal is then split: one is input
into a FADC (CAEN V1724, 100MHz bandwidth) while the other is input
into a discriminator to generate trigger control. Meanwhile, random
trigger and pulse modules are used to study efficiency feature etc.
The data from the FADC is transferred to PC through a
duplex optical fiber.
\begin{figure}[h]
\vspace*{0.2cm}\caption{~~Simplified Schematic Diagram of 20g-ULEGe DAQ setup.\\~} \label{fig18}
\hspace*{-0.5cm}\begin{minipage}[h]{\textwidth}
\includegraphics[scale=1.1]{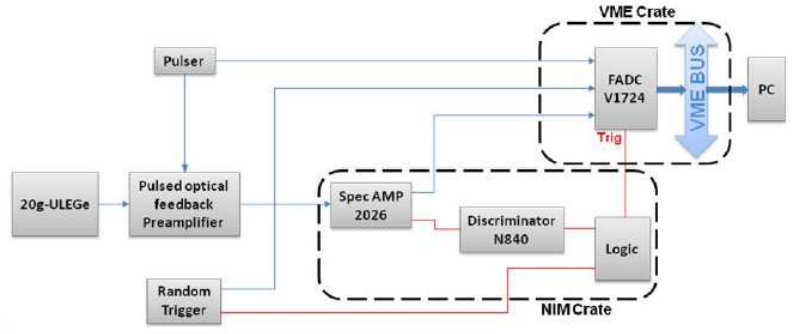}
\end{minipage}
\end{figure}
\subsubsection{1kg-PPCGe}

The Germanium detector 1kg-PPCGe(P-type) manufactured by Canberra,
France, was transported to the CJPL in 2011. This 1kg-PPCGe
detector, using new point contact technology, has a larger mass: 1kg
(single crystal). The crystal cylinder has an N+ contact on the
outer surface, and a tiny P+ contact stands as the central
electrode. The diameter and depth of the P+ contact are in order of
1 mm. The small diameter reduces the capacitance (order of 1pF) of
the detector, and readily improves the intrinsic noise standard
\cite{luke1989,barbeau}. The cryostat is designed fully-closed
within 1.5 mm thick copper layer, and there is no thin polymer film
window.

The 1kg-PPCGe possesses two preamplifiers, as shown in
Fig.\ref{fig19}. The p-type contact signal is read out by a pulsed
optical feedback preamplifier, with a low noise EuriFET nearby,
while the n-type contact signal is read out by a resistive feedback
preamplifier. Both preamplifiers have two identical outputs: OUT T
and OUT E. All outputs are connected to the downstream modules which
have high input impedance.

For the second phase of CDEX-1 experiment, the DAQ set up for
running the 1kg-PPCGe  is simply  illustrated in Fig.\ref{fig20}.
Considering the low energy range we are interested in, a fast timing
amplifier (Canberra 2111) is utilized to amplify the preamplifier
signal, and then input into a faster FADC (CAEN V1721, 500MHz
bandwidth). The other preamplifier outputs are directly connected
into conventional spectroscopy amplifiers (Canberra 2026). The
signal from N-type contact is then feeded into the FADC (CAEN V1724,
100MHz bandwidth); while the one from P-type contact is split into
the same FADC and a discriminator for trigger control respectively.
Again, the random trigger and pulse modules are used to study
efficiency features and etc. The data are transferred into PC's
through a duplex optical fiber.
\begin{figure}[h]
\vspace*{0.2cm}\caption{~~The front-end electronics of 1kg-PPCGe in CDEX-1.\\~} \label{fig19}
\hspace*{-0.5cm}\begin{minipage}[h]{\textwidth}
\includegraphics[scale=1.1]{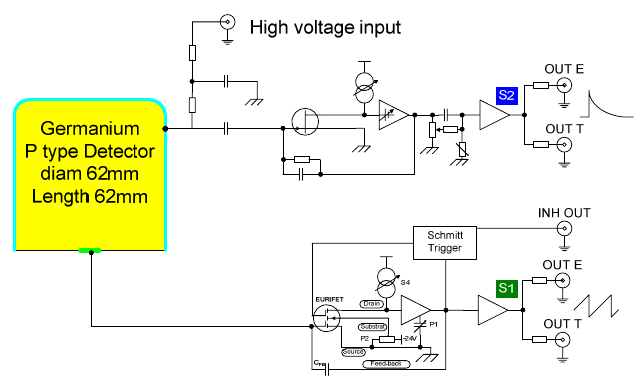}
\end{minipage}
\end{figure}
\begin{figure}[h]
\vspace*{0.2cm}\caption{~~Simplified Schematic Diagram of 1kg-PPCGe DAQ setup.\\~} \label{fig20}
\hspace*{-0.5cm}\begin{minipage}[h]{\textwidth}
\includegraphics[scale=1.1]{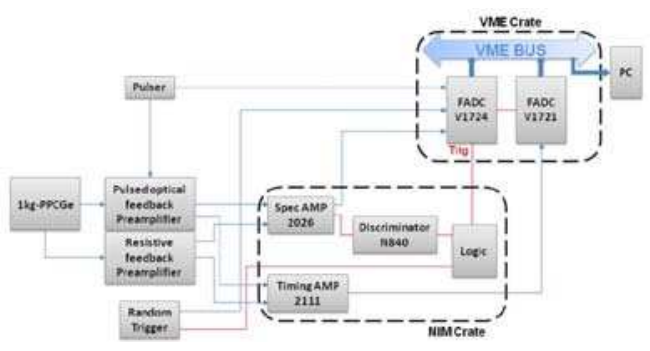}
\end{minipage}
\end{figure}

\subsubsection{Veto detector}

The passive shielding system has been well established which can
efficiently block the outside gamma ray coming inside, but cannot be
100$\%$. Besides, the shielding materials, even the HPGe detectors,
have radioactivity themselves at certain levels. If these rays
deposit energy in the HPGe detector, it would be hard to be
distinguished from the WIMP signals. Luckily, due to the extremely
low cross section of WIMPs colliding with ordinary matter, it is
almost impossible that WIMPs can deposit energy both in the HPGe
detectors and surrounding materials at the same time. So, the Veto
detectors are installed to surround the HPGe detector. By that
design, the output signal from HPGe is used as the trigger signal,
if there is also signal from surrounding Veto detectors during the
same time window, the detected event is not from collisions between
WIMPs and detector material and will be filtered. In addition, these
Veto detectors are also used for passive shielding. In summary, the
Veto detector should have following features:
\begin{itemize}
\item High Z and high density to achieve high veto efficiency and the high shielding ability;
\item Low internal radioactivity;
\item Low threshold to increase the veto efficiency.
\end{itemize}

In CDEX-1, CsI(Tl) or NaI(Tl) scintillation detectors are proposed
for the Veto detector. Except the features mentioned above, the
CsI(Tl) or NaI(Tl) scintillation detectors have capacity of carrying
the pulse shape discrimination (PSD), robust (reasonably soft and
malleable, less brittle and deliquescent for CsI(TI)) and easy to
make a large volume. Fig.\ref{fig21} shows the 18 CsI(Tl) crystal
array of veto detectors arranged around the cryostat, and each one
is read out by a PMT through a light guide. The advantages of this
design are: 1) the transportation length of scintillation light is
short,  good for light collection and lowers the threshold; 2) easy
for extension to accommodate larger PCGe detector; 3) the CsI(Tl)
crystal bar is easy to produce.
\begin{figure}[h]
\vspace*{0.2cm}\caption{~~The schematic diagram of CsI(Tl) veto detector system.\\~} \label{fig21}
\hspace*{-0.5cm}\begin{minipage}[h]{\textwidth}
\includegraphics[scale=1.1]{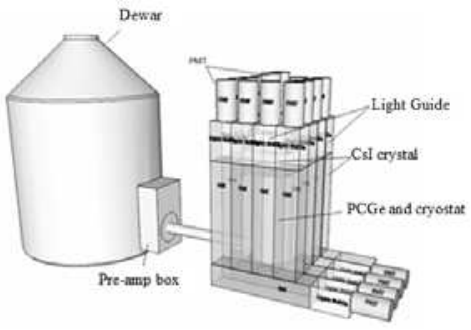}
\end{minipage}
\end{figure}

\subsection{CDEX-10}

Two detectors of CDEX-1, the 20g ULE-HPGe and 1kg-PPCGe, are running
at present in CJPL. The data analysis framework has also been set up
for the forthcoming data analysis. The CDEX-1 experiment is just the
first stage for using low energy threshold Ge detector to directly
detect dark matter. The CDEX-1 experiment will provide more detailed
information about the Ge detector performance and the background in
the active and passive shielding systems of the CDEX in CJPL; at the
same time, the preliminary physical results on dark matter search
with these two Ge detectors will be given  by the CDEX Collaboration
soon.

According to the recent model-dependent estimation on the cross
sections of collision between WIMPs and nuclei, the detector target
mass should be up to ton-scale if the dark matter experiment tries
to "see" the dark matter, i.e. achieve statistically sufficient
event rates with relatively high sensitivity. So the CDEX
collaboration plans to directly detect dark matter with low energy
threshold Ge detector of ton-scale target mass. There are several
basic conditions which should be considered for the future ton-scale
Ge detector. First, the Ge detector module with low energy threshold
down to sub-keV scale and enhancing the target mass of the total Ge
detector can be realized by increasing the number of the Ge detector
modules. The second one is that the Ge detector has to include a
cooling system to guarantee a low working temperature for the Ge
detector. The third one is that the Ge detector should possess a
veto detector serving as an active shielding against the background
contribution coming from the Ge detector itself and the material
near the Ge detector.

With this consideration, the whole detector system  is composed of
two parts: the Ge detector which can be up to ton-scale, is equipped
with 1kg mass Point-Contact Ge detector modules; another part is the
Liquid Argon veto detector serving as the cooling system and active
shielding. In order to test the conceptual design of this detector
and learn more details about the technology and gain experience on
the Ge detector  and LAr veto system, the CDEX Collaboration has
designed and will run a 10kg-scale Ge detector with LAr cooling and
veto system firstly.

Based on the studies of the PCGe detector, a new idea about the
active shielding system of the PCGe has emerged. The PCGe detector
will reach a ton-scale in the future in order to be more sensitive
to dark matter detection. In that scenario, the active shielding
system with the solid scintillators would be very difficult to
enclose the larger PCGe array system, while keeping the PCGe
detector to be cooled down to the liquid nitrogen temperature. So a
new type of active shielding technology for large-mass array PCGe
system has to be invented. The Liquid Argon (LAr) is a good
candidate. The temperature of LAr is suitable for the Ge detector
and LAr is also a kind of scintillator and can serve as an active
shielding system while the Ge detector is immersed into it.

The CDEX Collaboration has completed the physical and structure
simulation study on the CDEX-10 detector system, and also the
conceptual design and related studies on the Ge detector, LAr veto
detector, LAr cooling system and passive shielding system.

The structure of the whole CDEX-10 detector and the shielding system
are shown as follows. The interior of a 1 m thick (red) polyethylene
serves as a neutron shielding room. This neutron shielding room had
been constructed as the CJPL was under construction and now we have
started to carry out several experiments inside this polyethylene
room. The blue layer is lead shielding against the outside gamma
background. The yellow layer is Oxygen Free High purity Copper
(OFHC) which shields out the residual gamma background passing
through the lead layer and also shields the internal background
radiated by the lead shielding. Inside the OFHC layer there is the
LAr veto detector system which serves as both active and passive
shielding. Three 0.5 Kg or 1 kg PCGe detectors are encapsulated into
ultra-pure copper or aluminum tubes which are highly evacuated. One
block of the OFHC (Blue box) is mounted to shield the PCGe detector
from noise produced by the electronics in the tube. Several
encapsulated tubes with three PCGe detectors are immersed in the LAr
vessel for cooling and active shielding. The Total mass of the PCGe
detector is about 10 Kg. Part of the energy is deposited and the
scintillation light is produced when gamma or beta ray enters the
LAr. The scintillation light is read out by the photomultiplier
tubes£¨PMT£©and then the signal can serve as a veto signal.

The Ge detector tubes  on the 1 kg p-type point-contact Germanium
detector has been manufactured by the Caberra Company according on
the CDEX requirement and design. The 1 kg p-type PCGe detector  for
dark matter search experiment is also the biggest one in the world
till now. The 1 Kg-PCGe detector has been taking data after its
performance test.

In this CDEX-10 detector, the scintillation light is read out by the
PMT on the top of the LAr volume. The number of PMT is determined by
the detail of the physical design and the following plot gives the
layout plot of the PMT deployment.

The design of the cooling and active shielding systems for CDEX-10
LAr has been completed. From the physical point of view, the main
requirements for the LAr cooling and active shielding systems are to
keep the LAr temperature uniform and stable within a small
temperature region to alleviate the temperature dependence of the
performances of the PCGe detector, and to prevent  formation of gas
bubble and concomitant tiny vibration of LAr which may induce extra
noise in the Ge detector. A number of methods for satisfying such
stringent restrictions are considered to maintain constant liquid
level and control the generation of gas bubble in liquid argon under
zero boil-off conditions for long periods of time. Large convective
motions and pool-boiling are avoided by thermally optimized
cryogenic systems to reduce environmental heat leaks to the
low-temperature cryogen, especially, an actively-cooled LAr shield
that surrounds the cryostat is used against  heat radiation.

\subsection{Electronics (FADC, DAQ)}

The CDEX electronics includes three parts:  Electronic, Trigger
System and  Data Acquisition System where:
\begin{itemize}
\item The electronics includes the front-end amplifier, the main amplifier, the flash analog/digital converter, as well as HV power supply and the slow-control electronics etc.
\item The trigger system contains the trigger system and the clock distribution system.
\item The data acquisition system contains the data read-out electronics, the slow-control electronics and related data server and memory, communication, display and HMI etc.
\end{itemize}

\subsubsection{The Current  Design of Electronics}

Considering the future development, a set of multiple-channel
electronics for the CDEX detectors has been designed.
Fig.\ref{fig22} shows the design architecture of the electronics for
our CDEX. The mode is prevailing for both electronics and other
parts. So, in the following discussion, a thorough description about
the ongoing electronics is made first, and then we will go on
describing the next generation of electronics which is under research
$\&$ development, as well as the concerned engineering problems.
 \begin{figure}[h]
\vspace*{0.2cm}\caption{~~The ongoing design architecture of electronics: CDEX.\\~} \label{fig22}
\hspace*{-0.5cm}\begin{minipage}[h]{\textwidth}
\includegraphics[scale=0.4]{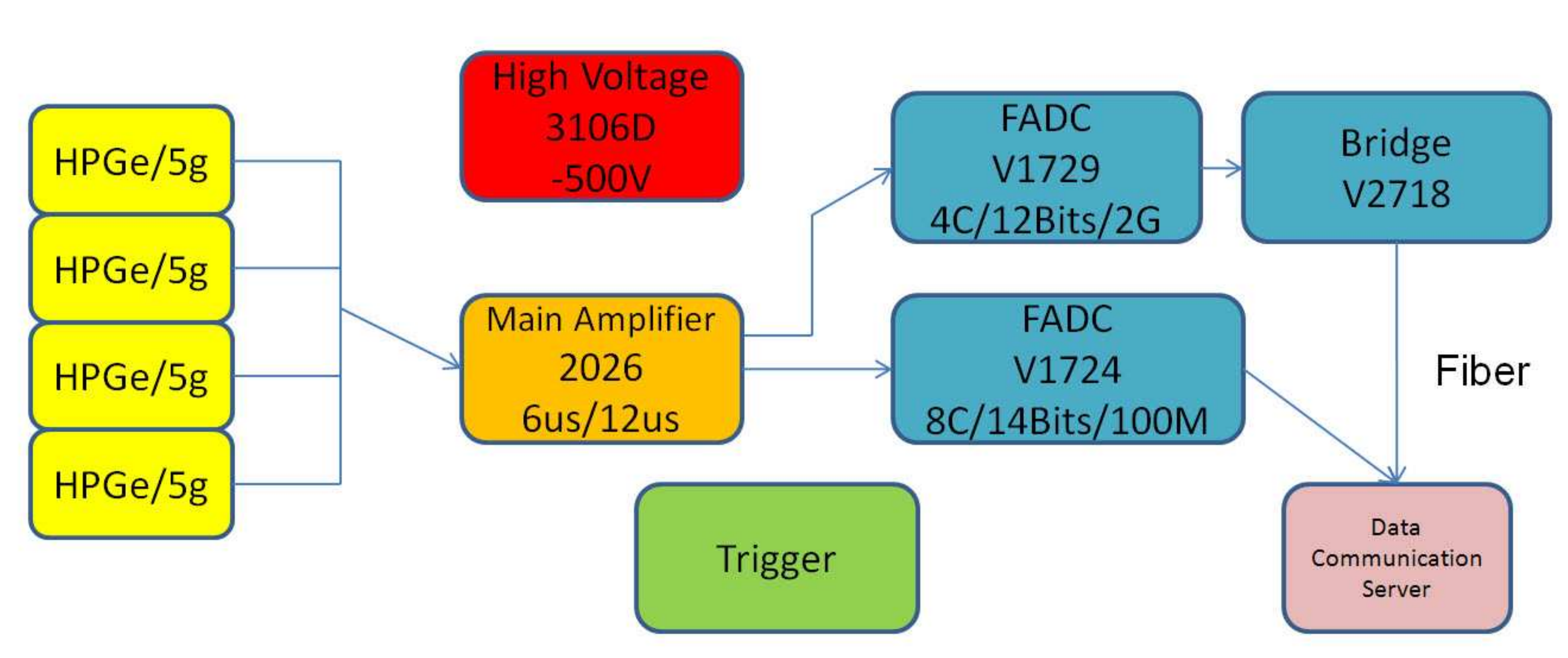}
\end{minipage}
\end{figure}

\subsubsection{The CDEX-1 and CDEX-10 Electronics Architectures}

It is forecasted that the  signal-readout channels of the
larger-scale CDEX detector will be 100$\sim$1000 in the future. It
is necessary to research and develop an  electronic system which
will be used for both HPGe  and veto detectors. The whole electronic
system includes the front-end amplifier, main amplifier, FADC and
data readout etc. Now the electronic group is focusing their efforts
on the design and plan of each parts of the electronic system as
described in the following.

\begin{itemize}
\item FADC:\\ According to the physical demand, there are two different
kinds of the FADC (Flash Analog to Digital Convertor)
electronics needed for reading out the pulse shape from the HPGe
detector or Veto detector: one is 100MHz, 14 Bits, 16-channel
FADC/GE electronic plug-in with standard dimensions of VME
plugin; another is 1GHz, 12Bits, 4-channel FADC/AC electronic
plugin with standard dimensions of VME plug-in. The FADC/GE
electronic plug-in is used to read out data on the
slow-shaping-time signal of the HPGe, and FADC/AC electronic
plug-in is used to read out the fast-shaping-time signal from
either HPGe or  photomultiplier tube (PMT) of the veto detector.

For convenience and to satisfy different physical demands, the
design of the FADC system adopts a new concept which is also
being applied in many labs of CERN. The structure uses the same
VME board (Main Board) and different FADC.  The experiments
facilities are realized by using different FADC Mezzanine Cards.
Different FADC mezzanine cards are made of FADC chips with
different performance indexes and numbers, the channels are
connected to the board via the high-speed plug-in. Then the same
board is used to conduct data cache, processing and triggering
judgment, and then the signals are read out via the standard VME
bus of the front-end fiber-optic interface of the back board.

\item The Readout Electronics:\\The CDEX electronic readout system is
integrated on the Wukong board which is an electronic product
developed by the researchers of Tsinghua University, and the
module RAIN200A is used to read out the data of each Wukong
Board. The module RAIN200A is based on PowerPC predecessor,
where Linux2.6.X is running to realize the readout data from
FADC to Ethernet, with a bandwidth at least 95Mbps.

The RAIN200A has the following characteristics: The core
predecessor is the 32-bit car-grade MPC5125 made by Free scale,
with frequency of 400MHz; the core board is installed with a
256MB DDR2 industrial-grade memory and 4GB NAND Flash; the core
board is supplied with 100Mbps Ethernet interface and USB 2.0
High-Speed interface; the system runs Linux 2.6.29 kernel with
RT patch.

\item DAQ:\\The CDEX DAQ includes the electronic part acquiring
data from the electronic units, the online data judgment,
display and memory as well as the off-line data analysis etc.

Because the electronic unit uses the Ethernet and TCP/IP as the
interface of data readout, DAQ needs merely to acquire data per
the standard protocol TCP/IP from the electronic unit, so that
our scheme is very convenient. The architecture of the entire DAQ
system is also an Ethernet-based exchange one and so the
standard commercial switches, routers and Ethernet cards can be
used to acquire data for the DAQ. Fig.\ref{fig23} gives the
Wukong FADC/DAQ system for the CDEX-1 and CDEX-10.
 \begin{figure}[h]
\vspace*{0.2cm}\caption{~~The Wukong FADC/DAQ system for CDEX-1 and CDEX-10.\\~} \label{fig23}
\hspace*{-0.5cm}\begin{minipage}[h]{\textwidth}
\includegraphics[scale=1.1]{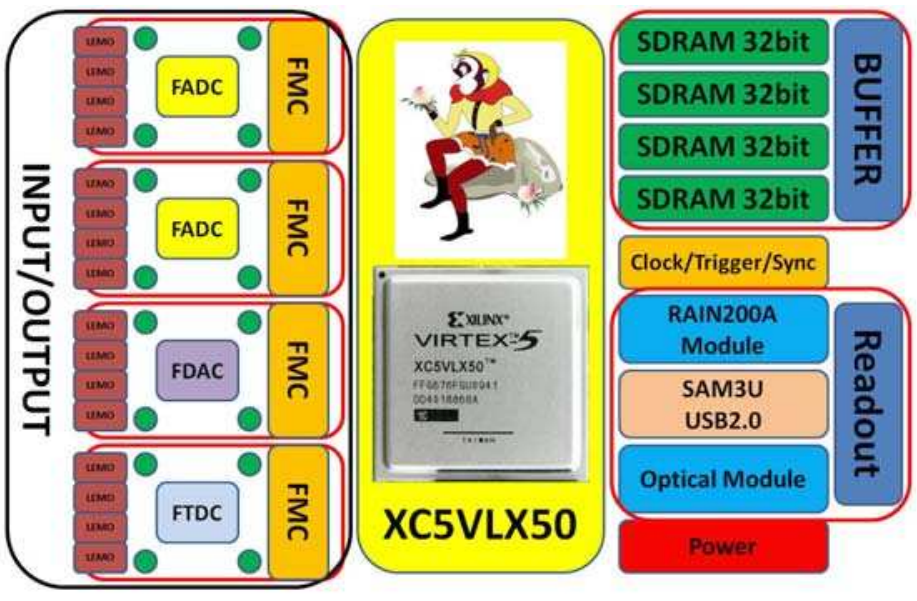}
\end{minipage}
\end{figure}
\end{itemize}
\section{Detector performance}

\subsection{Physical requirement}

As a detector used for dark matter detection, the most
important requirement for it is its low background level. The CDEX is
to detect WIMPs with a high purity germanium detector,
especially focusing on WIMPs of low masses ($<$10GeV). This imposes
another important requirement of energy threshold to the detector.
The CDEX Collaboration adopts the Point-Contact germanium detectors
(PCGe) for the low mass WIMP search which can realize a sub-keV
energy threshold with kg-size mass. The ton-scale detector system
of the CDEX in the future will be realized based on this kind of
kg-size PCGe detector. The PCGe detector has been developed from the
conventional Ultra Low Energy Threshold germanium detector. Many
efforts have been contributed to optimize the application of the
PCGe in dark matter searches.
\begin{itemize}
\item Pulse shape analysis of near noise-edge events extends the physics scope\\
the PCGe detector can provide ultra-low energy threshold down to
less than 500eV. According to the theories on dark matter, the
differential event rate of observing the recoiled nucleus
scattered off by incident WIMPs exponentially increases with the
decrease of its energy. So the pulse shape analysis of near
noise-edge events is a very important task for the CDEX
experiment to get lower energy threshold of the PCGe detector.
\item Pulse shape analysis of surface versus bulk events to characterize an important background channel\\
The location of an event where it is detected is another
important parameter for background discrimination. For a p-type
PCGe detector, due to the outside N+ or P+ conjunction, there
exists a thin layer which is called the "dead" layer. In fact, the
thin layer is not really "dead", the events emerging in this thin layer
can also give a different pulse shape from those bulk events
which are produced in the interior part of the PCGe detector. So
the pulse shape discrimination methods should be developed for
distinguishing the surface events from the bulk events.
\item Sub-keV background clarification and suppression\\
For the PCGe detector with a sub-keV energy threshold, it
provides us a new energy window where so far no researches have
ever covered this energy region due to relatively high energy
threshold of those detectors. The CDEX experiment should first
clarify the source and types of the background  in the sub-keV
energy region. Based on the present knowledge, a method for the
background discrimination and suppression will be developed.
\item Fabrication of advanced electronics for Ge detectors\\
The pulse shape discrimination methods have been developed for
suppressing  noise, especially near the energy
threshold, and discrimination of the surface events from bulk
events. All these methods to obtain lower energy threshold and
low background level are based on the performance of the
front-end electronics of the PCGe detector. So another important
requirement to the PCGe detector is the fabrication of advanced
electronics for Ge detector which provides relatively low noise
level.
\end{itemize}

\subsection{Performance of CDEX-1}

Since our first testing run of 20g-ULEGe in November of 2010, we
have obtained some data during both commissioning period and formal
data taking period. Data analysis is being undertaken for
determining the final spectra, and the properties of the 20g-ULEGe
in this new CJPL environment have been known.

The Commissioning period data of 1kg-PPCGe have also been acquired.
The properties of this detector are shown in the following
subsections.

\subsubsection{Linearity calibration of the detector 20g-ULEGe and 1kg-PPCGe}

Because of its small volume, the detection efficiency of 20g-ULEGe
is relatively low to get prominent peaks for calibration, even
though the calibration period was prolonged to several months.
Withal, there is no source available for further test in the CJPL at
present. An X-ray tube \cite{amtek}, which can generate X-ray of up
to about 30 keV, is applied as a substitution of the source. When
the X-ray generated by the tube hits the mixed powders of titanium
dioxide (TiO2) and potassium permanganate (KMnO4) the characteristic
X-rays can be used for the high gain (low energy range) calibration.
The surrounding copper shielding contributes characteristic X-rays
as well. Thanks to the thin carbon composite window of 0.6mm,
overwhelming amount of the low energy X-rays are able to penetrate
and deposit energies in the crystals. Sketch of the calibration for
the 20g-ULEGe is illustrated in Fig.\ref{fig24}.
 \begin{figure}[h]
\vspace*{0.2cm}\caption{~~Sketch of calibration for 20g-ULEGe by X-ray tube.\\~} \label{fig24}
\hspace*{-0.5cm}\begin{minipage}[h]{\textwidth}
\includegraphics[scale=1.1]{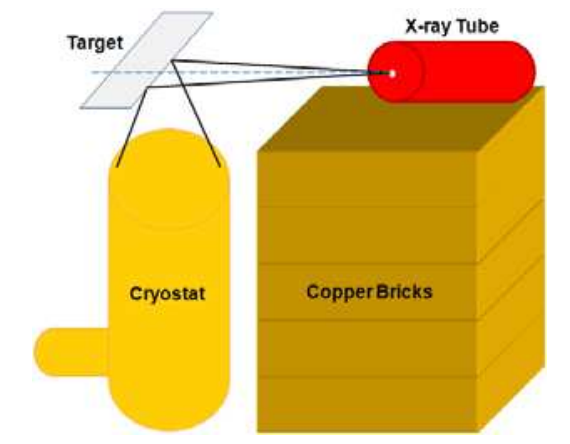}
\end{minipage}
\end{figure}

The calibration result of the 20g-ULEGe are shown in
Fig.\ref{fig25}, with each sub-figure in accord with a 5 g
sub-detector. In this scheme, characteristic X-rays of titanium,
manganese and copper, coupled with random trigger \cite{tek} (zero
point) are used (Table.\ref{Table1}). From the fitting results of
the seven points, we can observe a good linearity in the low energy
range below 10 keV. Moreover, the region above 10 keV can  be
studied with available radiation sources.
 \begin{figure}[h]
\vspace*{0.2cm}\caption{~~Calibration result of 20g-ULEGe.\\~} \label{fig25}
\hspace*{-0.5cm}\begin{minipage}[h]{\textwidth}
\includegraphics[scale=0.8]{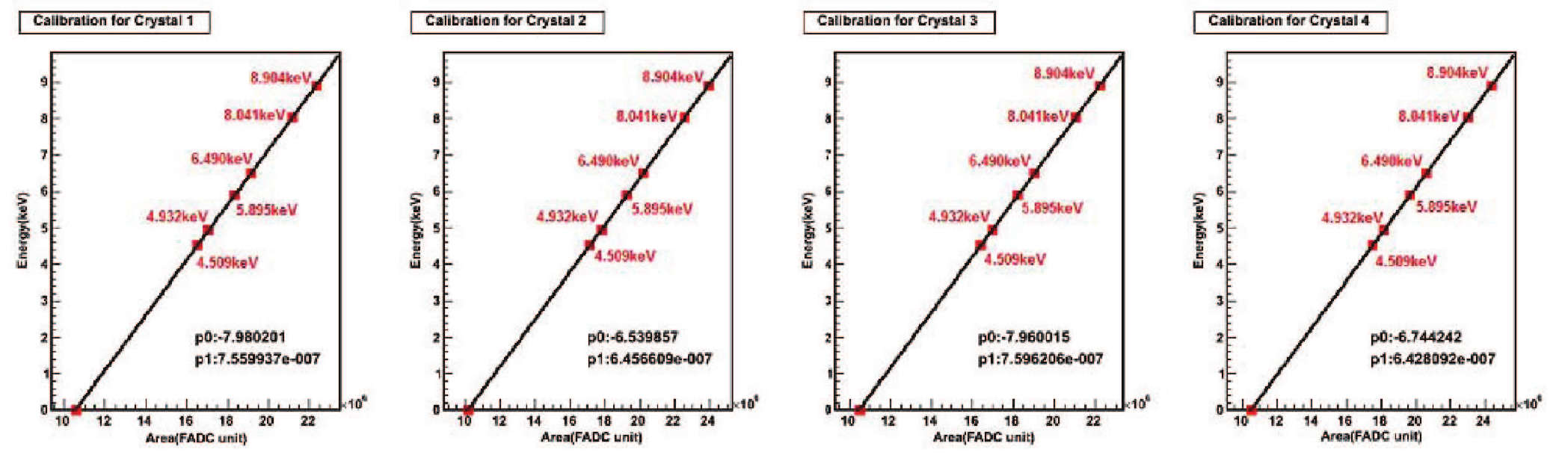}
\end{minipage}
\end{figure}
\begin{table}
\large\centering
\caption{Sources for calibration of 20g-ULEGe}\label{Table1}
\begin{tabular}{c|c|c|c|c|c|c|c}
\hline\hline
\multicolumn{1}{c|}{\multirow {1}{*}{Source}}&\multicolumn{1}{|c|}{\multirow {1}{*}{Random trigger}}&\multicolumn{2}{|c|}{Ti}&\multicolumn{2}{|c|}{Mn}&\multicolumn{2}{|c}{Cu}\\ \cline{1-8}
\multicolumn{1}{c|}{Energy (keV)}&\multicolumn{1}{|c|}{0}&\multicolumn{1}{|c|}
{4.509} & \multicolumn{1}{|c|}{4.932} & \multicolumn{1}{|c|}{5.895} & \multicolumn{1}{|c|}{6.490} & \multicolumn{1}{|c|}{8.041} & \multicolumn{1}{|c}{8.904}\\
\hline\hline
\end{tabular}
\end{table}

The 1kg-PPCGe does not have a carbon composite window, but a large
volume and mass. So its detection efficiency is high enough to carry
out a calibration with a few days' background data. The result of
one data set is shown in Fig.\ref{fig26}, with 3 sub-figures
corresponding to different gains. The period is 21.9 days and lots
of background peaks are visible. Besides the zero point by random
trigger, we respectively use 4-8 points to complete the calibration
fitting  for different gains (Table.\ref{Table2}). The results also
display a good linearity of the 1kg-PPCGe.
 \begin{figure}[h]
\vspace*{0.2cm}\caption{~~Calibration result of 1kg-PPCGe.\\Left: High Gain. Middle: Mediate Gain. Right: Low Gain.\\~} \label{fig26}
\hspace*{-0.5cm}\begin{minipage}[h]{\textwidth}
\includegraphics[scale=1.1]{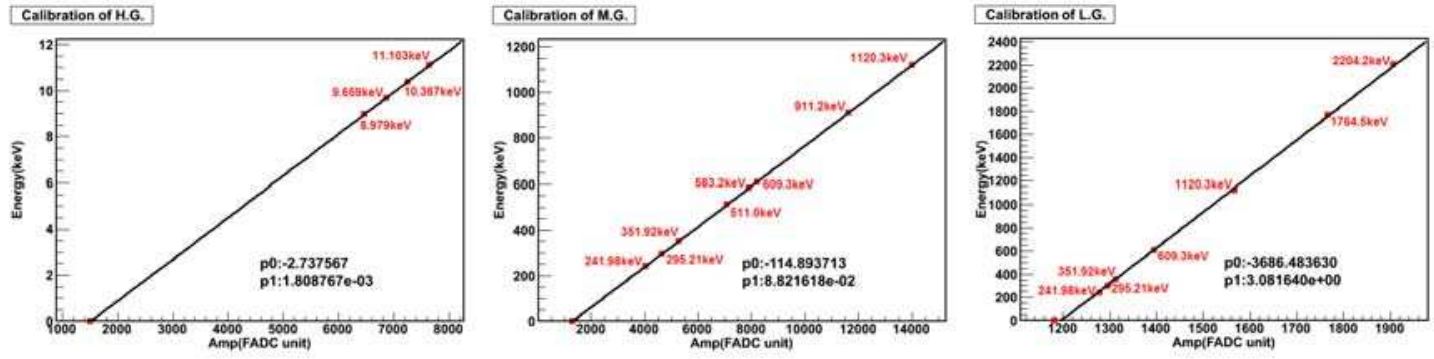}
\end{minipage}
\end{figure}
\begin{table}
\large\centering
\caption{Sources for calibration of 1kg-PPCGe}\label{Table2}
\begin{tabular}{c|c|c|c|c|c}
\hline\hline
\multicolumn{2}{c|}{High Gain}&\multicolumn{2}{|c|}{Mediate Gain}&\multicolumn{2}{|c}{Low Gain}\\ \cline{1-6}
\multicolumn{1}{c|}{Source}&\multicolumn{1}{|c|}{Energy (keV)}&\multicolumn{1}{|c|}
{Source} & \multicolumn{1}{|c|}{Energy (keV)} & \multicolumn{1}{|c|}{Source} & \multicolumn{1}{|c}{Energy (keV)}\\ \hline
\multicolumn{1}{c|}{$\mathrm{^{65}Zn}$}&\multicolumn{1}{|c|}{8.979}&\multicolumn{1}{|c|}
{$\mathrm{^{214}Pb}$} & \multicolumn{1}{|c|}{214.98} & \multicolumn{1}{|c|}{$\mathrm{^{214}Pb}$} & \multicolumn{1}{|c}{214.98}\\ \hline
\multicolumn{1}{c|}{$\mathrm{^{68}Ga}$}&\multicolumn{1}{|c|}{9.659}&\multicolumn{1}{|c|}
{$\mathrm{^{214}Pb}$} & \multicolumn{1}{|c|}{295.21} & \multicolumn{1}{|c|}{$\mathrm{^{214}Pb}$} & \multicolumn{1}{|c}{295.21}\\ \hline
\multicolumn{1}{c|}{$\mathrm{^{68}Ge}$}&\multicolumn{1}{|c|}{10.367}&\multicolumn{1}{|c|}
{$\mathrm{^{214}Pb}$} & \multicolumn{1}{|c|}{351.92} & \multicolumn{1}{|c|}{$\mathrm{^{214}Pb}$} & \multicolumn{1}{|c}{351.92}\\ \hline
\multicolumn{1}{c|}{$\mathrm{^{73,74}As}$}&\multicolumn{1}{|c|}{11.103}&\multicolumn{1}{|c|}
{$\mathrm{Ann}$} & \multicolumn{1}{|c|}{511.0} & \multicolumn{1}{|c|}{$\mathrm{^{214}Bi}$} & \multicolumn{1}{|c}{609.3}\\ \hline
\multicolumn{1}{c|}{$-$}&\multicolumn{1}{|c|}{-}&\multicolumn{1}{|c|}
{$\mathrm{^{208}Tl}$} & \multicolumn{1}{|c|}{583.2} & \multicolumn{1}{|c|}{$\mathrm{^{214}Bi}$} & \multicolumn{1}{|c}{1120.3}\\ \hline
\multicolumn{1}{c|}{$-$}&\multicolumn{1}{|c|}{-}&\multicolumn{1}{|c|}
{$\mathrm{^{214}Bi}$} & \multicolumn{1}{|c|}{609.3} & \multicolumn{1}{|c|}{$\mathrm{^{214}Bi}$} & \multicolumn{1}{|c}{1764.5}\\ \hline
\multicolumn{1}{c|}{$-$}&\multicolumn{1}{|c|}{-}&\multicolumn{1}{|c|}
{$\mathrm{^{228}Ac}$} & \multicolumn{1}{|c|}{911.2} & \multicolumn{1}{|c|}{$\mathrm{^{214}Bi}$} & \multicolumn{1}{|c}{2204.2}\\ \hline
\multicolumn{1}{c|}{$-$}&\multicolumn{1}{|c|}{-}&\multicolumn{1}{|c|}
{$\mathrm{^{214}Bi}$} & \multicolumn{1}{|c|}{1120.3} & \multicolumn{1}{|c|}{-} & \multicolumn{1}{|c}{-}\\ \hline\hline
\end{tabular}
\end{table}

\subsubsection{Energy resolution of detectors 20g-ULEGe and 1kg-PPCGe}

With the calibration data, the energy spectra as well as the
properties of energy resolution have been obtained.

The calibrated spectra of the 20g-ULEGe are plotted in
Fig.\ref{fig27}. FWHMs (Full Width at Half Maximum, fitted to a
Gaussian function) are about 200 eV at 5 keV and about 300 eV at 9
keV. The 4 sub-detectors are very close to each other in all
technical indices and practical performance. Besides the statistical
limit, the X-ray tube which has a wide emission solid angel and
notable energy dispersion, also degenerate the FWHM. So the energy
resolution of the system is expected to be better than the
predecessor products.
 \begin{figure}[h]
\vspace*{0.2cm}\caption{~~Calibrated energy spectra of 20g-ULEGe.\\~} \label{fig27}
\hspace*{-0.5cm}\begin{minipage}[h]{\textwidth}
\includegraphics[scale=0.8]{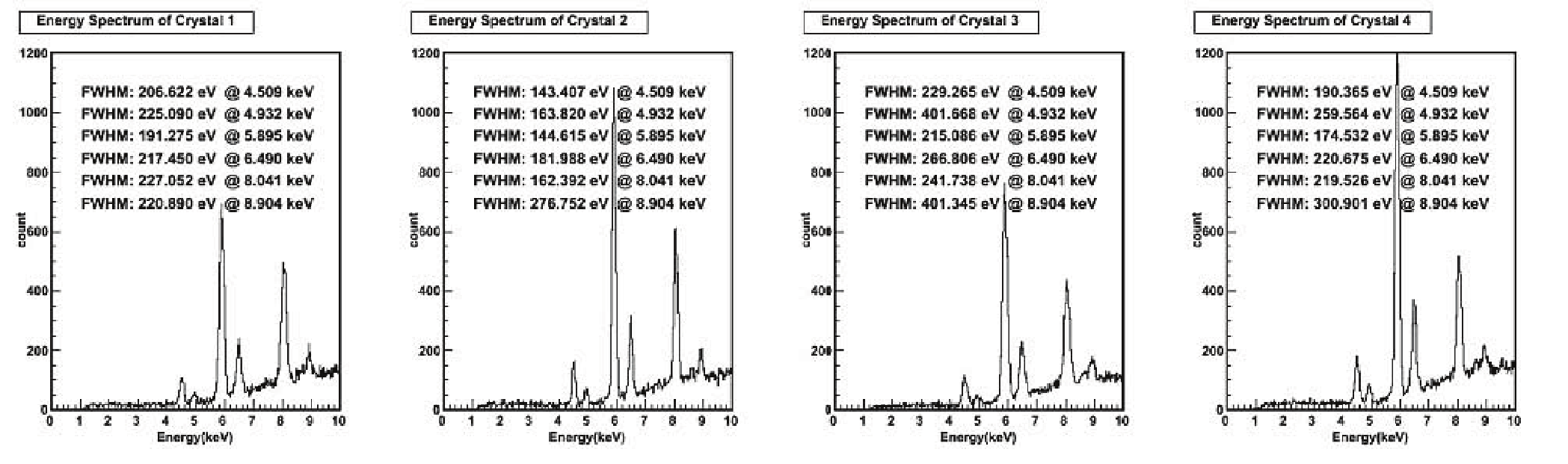}
\end{minipage}
\end{figure}

The calibrated spectra of 1kg-PPCGe are plotted in Fig.\ref{fig28}.
The FWHMs are about 200 eV at 10.367 keV and about 3.525 keV at
1120.3 keV. For many other peaks, FWHMs are limited by statistical
errors, thus need to be improved with more data accumulation.
 \begin{figure}[h]
\vspace*{0.2cm}\caption{~~ Calibrated energy spectra of 1kg-PPCGe.~Left: High Gain. Middle: Mediate Gain. Right: Low Gain.\\~} \label{fig28}
\hspace*{-0.5cm}\begin{minipage}[h]{\textwidth}
\includegraphics[scale=0.8]{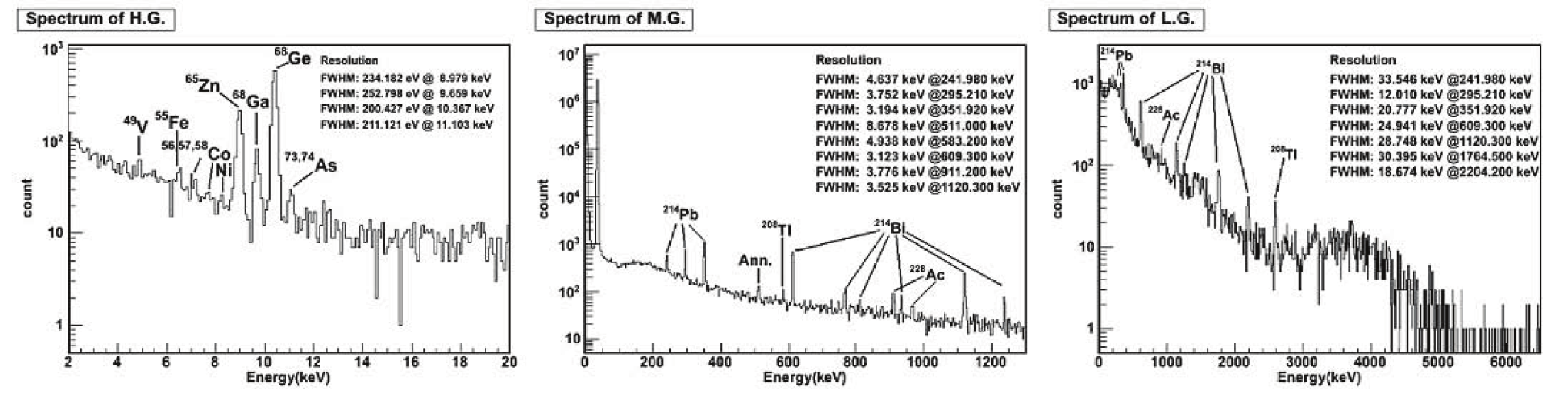}
\end{minipage}
\end{figure}

\subsubsection{Noise level of detector 20g-ULEGe and 1kg-PPCGe}

Electronic noise would crucially affect the detection threshold. To
study the noise level, events by random trigger tag are selected and
projected to energy spectra, which are then fitted with the Gaussian
function. The FWHMs of the distributions render less than 100 eV of
both 20g-ULEGe (crystal 2) and 1kg-PPCGe, which sustain the
threshold lower than 500 eV. (Fig.\ref{fig29},\ref{fig30})
 \begin{figure}[h]
\vspace*{0.2cm}\caption{~~Noise distribution of 20g-ULEGe by random trigger.\\~} \label{fig29}
\hspace*{-0.5cm}\begin{minipage}[h]{\textwidth}
\includegraphics[scale=1.1]{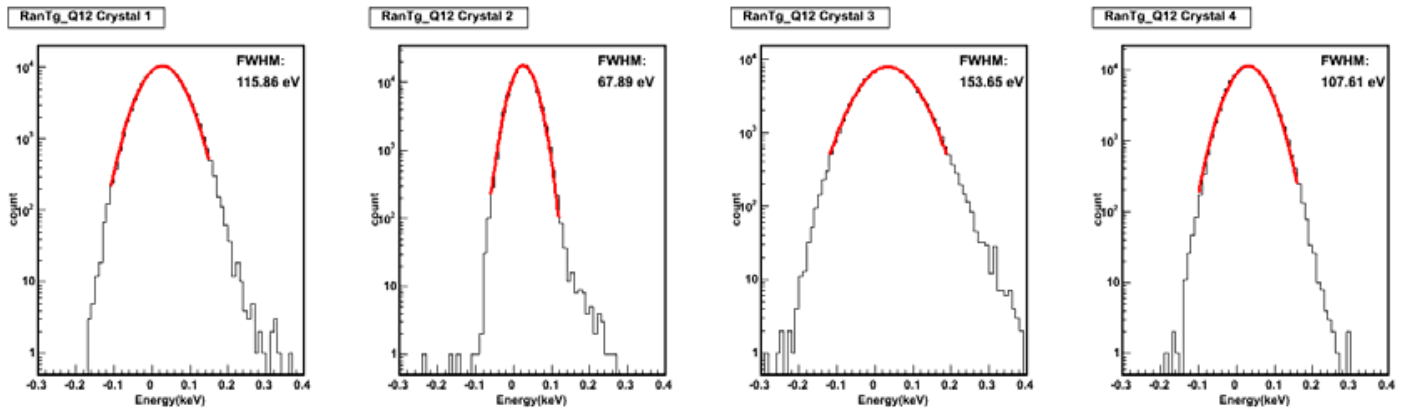}
\end{minipage}
\end{figure}
 \begin{figure}[h]
\vspace*{0.2cm}\caption{~~Noise distribution of 1kg-ULEGe by random trigger (High Gain).\\~} \label{fig30}
\hspace*{-0.5cm}\begin{minipage}[h]{\textwidth}
\includegraphics[scale=1.1]{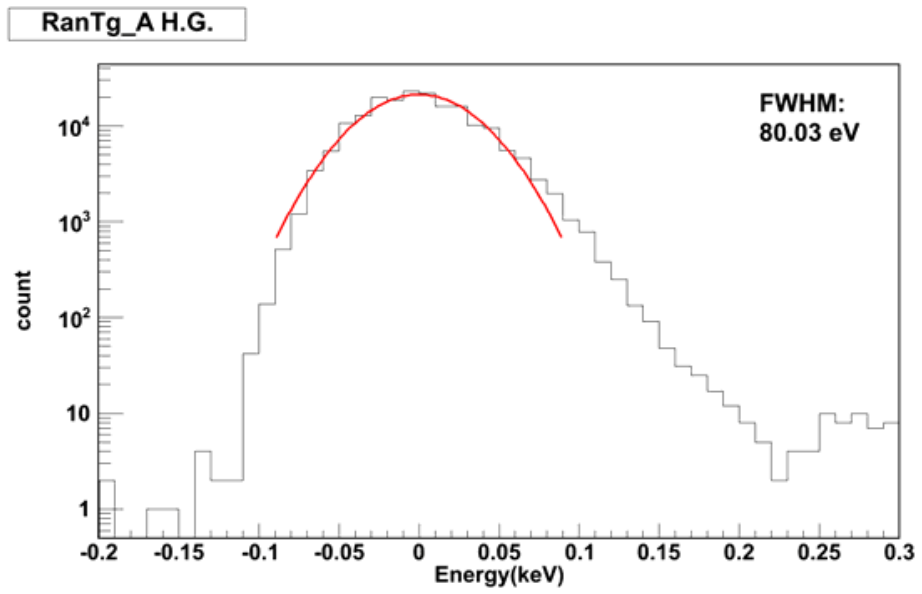}
\end{minipage}
\end{figure}

\subsubsection{Stability of the detector}

The validity of data which depict the system stability needs to be
checked before overall data analysis. Several parameters describing
the system behaviors are tracked over time, including triggering
rate, pedestal of pulse shape, noise level, et. al..
Figs.\ref{fig31},\ref{fig32} present the triggering rate status of
20g-ULEGe and 1kg-PPCGe, in which an average rate per hour is
counted. Immoderate parts, especially an anomalous discrepancy
deviating from the average level, should be carefully inspected and
discarded if they are  proved to be invalid.
 \begin{figure}[h]
\vspace*{0.2cm}\caption{~~Triggering rate status of 20g-ULEGe.\\~} \label{fig31}
\hspace*{-0.5cm}\begin{minipage}[h]{\textwidth}
\includegraphics[scale=1.1]{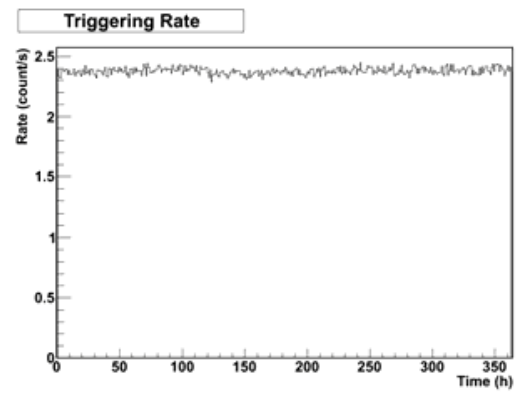}
\end{minipage}
\end{figure}
 \begin{figure}[h]
\vspace*{0.2cm}\caption{~~Triggering rate status of 1kg-PPCGe.\\~} \label{fig32}
\hspace*{-0.5cm}\begin{minipage}[h]{\textwidth}
\includegraphics[scale=1.1]{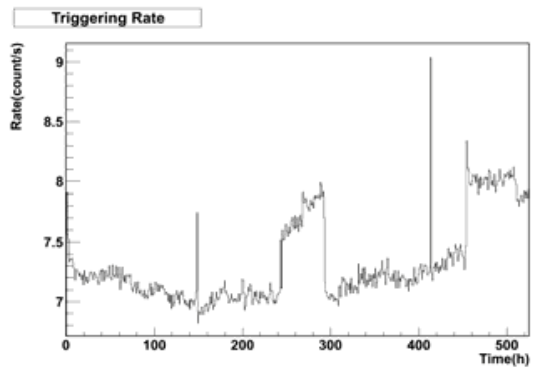}
\end{minipage}
\end{figure}

\subsubsection{Trigger efficiency}

Trigger efficiency is to be checked before data analysis is carried
on \cite{stlin2009}. It determines the survival fractions of the
events which pass the discriminator threshold. Specific pulse
generator which can supply small signal ranged at 0-2 keV can be used
to estimate the trigger efficiency at lower energy regions. Such a
kind of pulse generators is under development right now, and the
relevant test will be held soon.

\section{Data analysis}

\subsection{Process}

The data analysis is a very important step to understand the
energy spectra of the HPGe detectors. Usually there are three
sources contributing to the HPGe spectra: electronic noises and
microphonic interference, physical background events as well as dark
matter events. The purpose of the data analysis is first to
discriminate out the noise, physical background events from the
total energy spectra. Different methods need to be developed
according to the characteristics of different sources.

The signal shape of electronic noise and microphonic interference
are not in the Gaussian-like wave form in which the physical events
should be, then can be distinguished through some PSD (Pulse Shape
Discrimination) method. Various parameters, time-to-peak, rise time
and fall time, for instance, describing wave form of the signal can
be designed and used for differentiating in parameter plots. In
contrast to the anomalous wave form, analog to the Gaussian-like
pulse shape would be a criterion for identifying the type of the
sources in the calibration data.

In particular, electronic noise exerts a crucial impact on the threshold
and needs to be handled with great caution. Physical events, in
principle, follow the Gaussian distribution in 2-D energy-energy
parameter plot. So the threshold can be lowered in this parameter plot
than in a single energy spectrum plot. A noise edge cut, which
exists along the tangent of the Gaussian distribution, is then set
to lower the threshold. Since the cut is related to energy, the
efficiency-energy curve should be estimated by the pulse calibration
data or anti-Compton tagged data.

Usually in experiments of searching for dark matter one observes
recoiled nuclei, such events might be tangled with the physical
background caused by neutrons and electron recoil events. Several
experiments, such as CDMS\cite{cdms}, XENON100\cite{xenon},
CRESST-II\cite{cresst} et. al., use two distinct signals such as
scintillation and ionization to reject the electron recoil events
simultaneously. But it's not the case in the CDEX. The CDEX only
measures the energy deposit of ionization. Another difference is
that the CDEX needs to consider the quenching factor of energy
deposit for the recoiled nucleus.

As the pulses caused by the electron recoil and nucleus recoil do
not appear different, the normal PSD method would be less efficient
here, some other aspects need to be viewed. There are mainly two
ways to subtract the  background events. One is to statistically
eliminate the visible and invisible characteristic x-ray peaks;
while the other is to apply some algorithm to reduce surface events
occurring in the thin insensitive layer. In the low energy region
which we are interested in, visible characteristic peaks usually
come from the K-shell X-rays and the invisible ones are caused by
the L-shell X-rays' contribution, both of them have been analyzed
theoretically and experimentally in some details\cite{aalseth}. On
the other hand, the surface insensitive layer of the PPCGe detector
has a weaker electric field therefore can cause distortion of energy
deposition. Such a kind of events would have slow pulse shape from
preamplifier, but is degenerate with the electronic noise. The
Wavelet method \cite{marino}  can reduce noise by filtering and
shaping the signal sample.

With all the aforementioned processes, the final energy spectrum is
obtained. Based on selected physics and statistical method, the final
spectrum is then interpreted into physics result, namely the common exclusion curves.

\subsection{Strategy}

In order to measure the cross sections for low WIMP masses,
elaborate data analysis is essential to lower the threshold and
single out background events from raw data. The threshold
can be lowered by getting rid of noise events, and the non-WIMP
event rate can be suppressed by deducting the microphonic events as
well as physically identified non-WIMP events. Microphonic events,
coming from violent changes of environment or instability of the
system, usually have abnormal pulse shape. Non-WIMP events may
consist of multi-site events (events with several incident particles
coincidentally), surface events (events with energy deposited in the
dead layer of the surface), et. al.. To attain this goal, some
variables or parameters are defined, either event by event or
statistically. Fig.\ref{fig33} illustrates the notation of several
empirical main parameters. Ped, e.g., is calculated as the average
pedestal within the first 200 FADC time units.
 \begin{figure}[h]
\vspace*{0.2cm}\caption{~~Notation of several parameters.\\~} \label{fig33}
\hspace*{-0.5cm}\begin{minipage}[h]{\textwidth}
\includegraphics[scale=1.1]{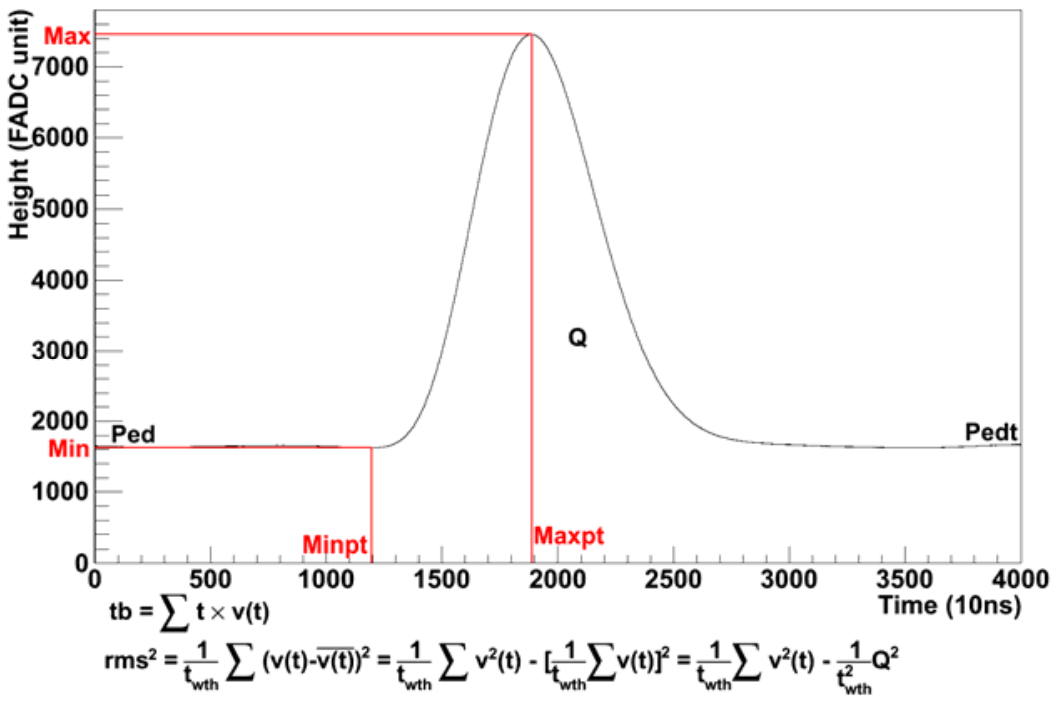}
\end{minipage}
\end{figure}

Some basic cuts for selection of pulse shapes, such as the stability
of pedestal, time-to-peak, as well as linearity between area and
amplitude, are applied firstly. After the pulse shape selection,
three different crucial cuts are utilized as described in the
following subsections. Then the derived final spectrum is used to
interpret the physics, commonly as making exclusion plots or
indicating islands in the parameter space.

Moreover, to identify the noise and microphonic events, reference
pulse shape of physics events are needed. Such physics events can be
selected within peaks of characteristic x-ray, either from
calibration data or from background measurement.

\subsubsection{PSD cut}

Noise and microphonic events having different pulse shapes from the
physics events, can be distinguished through the PSD (Pulse Shape
Discrimination) method. The parameters for each event pulse with
different shaping time are then considered, and the scattered plot
is shown in Fig.\ref{fig34}. The two parameters are the pulse
amplitude with shaping times of $\mathrm{6\mu s}$ and $\mathrm{12\mu
s}$ respectively. Red dots roughly along the diagonal stand for the
physics and noise events, while the black dots in the band near the
axes correspond to the interference events. So taking the straight
lines (blue-colored) as cuts, we are able to discard the microphonic
events.
 \begin{figure}[h]
\vspace*{0.2cm}\caption{~~Illustration of PSD cut \cite{stlin2009}.\\~} \label{fig34}
\hspace*{-0.5cm}\begin{minipage}[h]{\textwidth}
\includegraphics[scale=1.1]{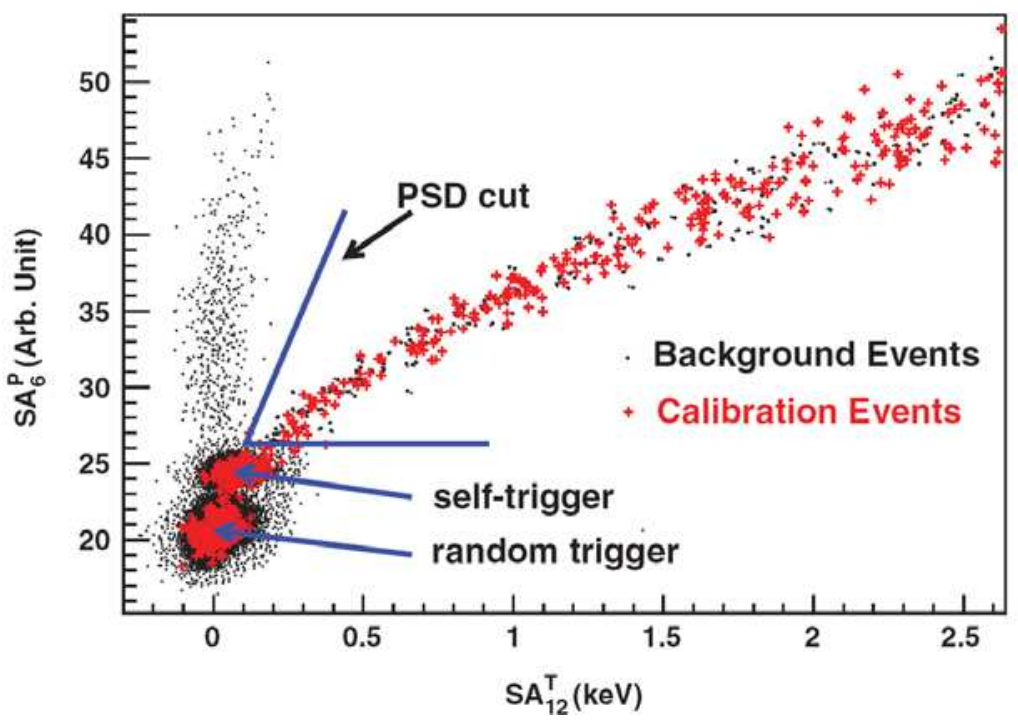}
\end{minipage}
\end{figure}

\subsubsection{Noise edge cut}

In principle, the pulse amplitude distribution of the noise events
is Gaussian. So in the 2D (pulse amplitude and pulse area) parameter
plot, the distribution seems like an ellipse (Fig.\ref{fig35}).
Whenever they are projected to either axis, the bulge would degrade
the noise level. Noise edge cut is then used to reduce the noise
events and hence lower the threshold. The key point is that since
this cut is related to the energy, the efficiency curve has to be
applied to make corrections.
 \begin{figure}[h]
\vspace*{0.2cm}\caption{~~Illustration of noise edge cut.\\~} \label{fig35}
\hspace*{-0.5cm}\begin{minipage}[h]{\textwidth}
\includegraphics[scale=1.1]{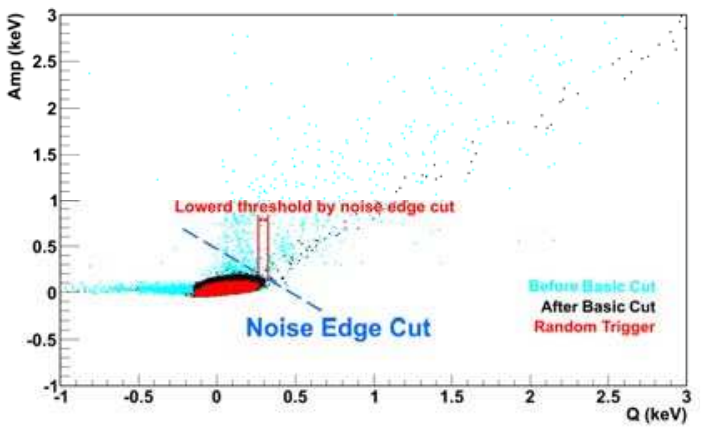}
\end{minipage}
\end{figure}

\subsubsection{Surface event cut}

Due to the manufacturing limitation, the N+-type surface layer of
the PPCGe is thick, in which the electric field is weaker than in
the crystal bulk so that the energy deposited in this layer cannot
be efficiently collected. This layer is then named as the dead layer
or insensitive layer, and the recorded events with energy deposited
in this layer are called as surface events. The surface events have
a longer charge-collection time and slower pulse shape, which may be
picked out by the time parameter cut. The rising time of the fast
pulse at the preamplifier  is chosen as the time parameter. However,
in the low energy range which we are highly interested in, the
signal-to-noise ratio becomes not high enough to finely deduce the
parameter. Green curve in Fig.\ref{fig36} shows a raw fast pulse.
Marino \cite{marino} suggested a method that one can use a wavelet
shrinkage to reduce noise. The result is shown as the black curve,
and the rise time of 10-90$\%$ of leading edge is then estimated.
 \begin{figure}[h]
\vspace*{0.2cm}\caption{~~Illustration of wavelet analysis \cite{marino}.\\~} \label{fig36}
\hspace*{-0.5cm}\begin{minipage}[h]{\textwidth}
\includegraphics[scale=1.1]{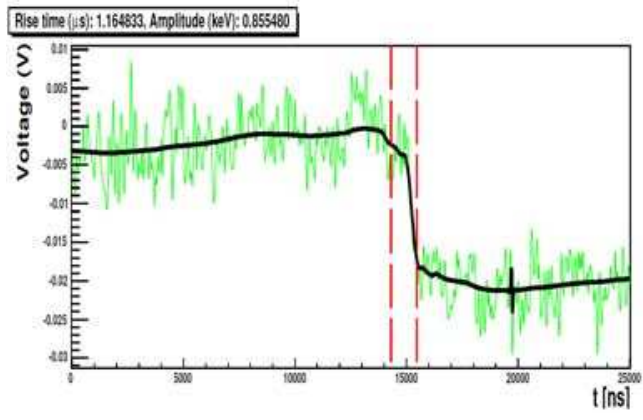}
\end{minipage}
\end{figure}

A typical tendency between the rising time and energy distribution is
presented in Fig.\ref{fig37}, where two gatherings are clearly
separated. The up part which has longer rising time corresponds to
the surface events, so that should be discarded by the red critical
line. The down part is then determined to be the final spectrum.
Efficiency correction is needed afterwards, which could be estimated
by source calibration data.
  \begin{figure}[h]
\vspace*{0.2cm}\caption{~~Illustration of the surface event cut \cite{wilkerson}.\\~} \label{fig37}
\hspace*{-0.5cm}\begin{minipage}[h]{\textwidth}
\includegraphics[scale=1.1]{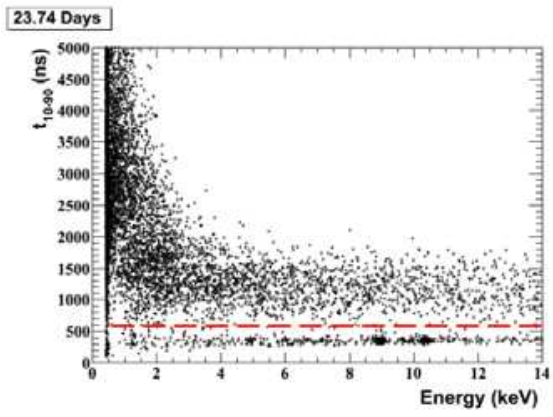}
\end{minipage}
\end{figure}

\section{The prospect}

\subsection{Physical target}

The CDEX Collaboration will realize the dark matter detection with
point-contact germanium arrays as the detector which is cooled down
and actively shielded by the liquid argon system. The most
interesting region for the CDEX experiment is the low WIMP mass
region which could be detected  by  the ultra-low energy threshold
PCGe detector. Based on the performances of the prototype 1 kg PCGe
detector which has run for several months in the shielding system at
CJPL, the CDEX collaboration  plans to focus on the direct detection
of dark matter particles with a minimum  mass less than 10 GeV. The Ge array
system will run in the CJPL and additional shielding system will be
included such as 1 m-thick polyethylene for neutron deceleration and
absorption, 20 cm lead and 20 cm Oxygen-Free High-Conductivity copper
for gamma shielding. A liquid argon cooling and active shielding
system will be adopted for the CDEX-10, and the PCGe detector will
be immersed into the liquid argon system. The passive and active
shielding systems  provide the Ge array system a relatively low
background circumstance, thus running experiments with low
background is expected. Except the shielding system, the effective
pulse shape discrimination (PSD) methods will also be developed to
get rid of the noise and background events from the raw data and pick up the real dark matter
events. These PSD methods  mainly focus on (1) the pulse shape
analysis of the events near noise-edge to enlarge the observable
physics range, (2) the pulse shape analysis of surface events, by
comparison with the bulk events one can characterize the important
background channel, (3) understanding the sub-keV background and
offering a  scheme to suppress it.

By  considering all possible effects in the design which might
affect the performances of the detector and carrying careful
dada analysis, we wish to achieve a level of 1cpkkd (count per
kilogram per keV per day) for the CDEX-1 stage and 0.1cpkkd for
CDEX-10 stage. The different exclusive curves corresponding to
different energy thresholds and background event rates are shown in
Fig.\ref{fig38}.
  \begin{figure}[h]
\vspace*{0.2cm}\caption{~~The different exclusive curves correspond to the different energy threshold and background event rate.\\~} \label{fig38}
\hspace*{-0.5cm}\begin{minipage}[h]{\textwidth}
\includegraphics[scale=1.1]{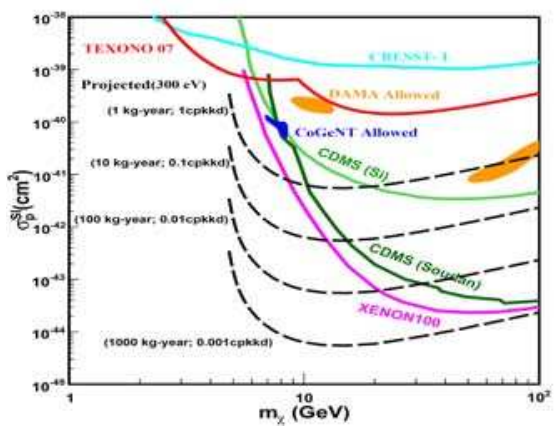}
\end{minipage}
\end{figure}

\subsection{CDEX-1T}

The ultimate goal of the CDEX Collaboration is to set up a ton-scale
mass Ge detector based on the PCGe detector and LAr active shielding
and cooling systems. The detailed design and technology which will
be developed and utilized are based on the experience and lessons
gained from the design and studies of the CDEX-10 detector. The
detector will be located in a 1 m-thick polyethylene shielding room
and the internal volume is covered by 20 cm lead and 10 cm
Oxygen-Free High-Conductivity copper. The threshold less than 400 eV
and background event rate of 0.001 count per keV per kilogram target
mass per day for the 1 T Ge mass are the main goal of the CDEX-1T
detector. The sensitivity for low mass dark matter will be down to
about $\mathrm{10^{-45}cm^2}$ in the region of the WIMP mass less
than 10 GeV if the 0.001cpkkd background level is achieved, as shown
in Fig.\ref{fig38}. The rough layout of the CDEX-1T detector is
shown in Fig.\ref{fig39}.
  \begin{figure}[h]
\vspace*{0.2cm}\caption{~~Layout of the CDEX-1T detector system.\\~} \label{fig39}
\hspace*{-0.5cm}\begin{minipage}[h]{\textwidth}
\includegraphics[scale=1.1]{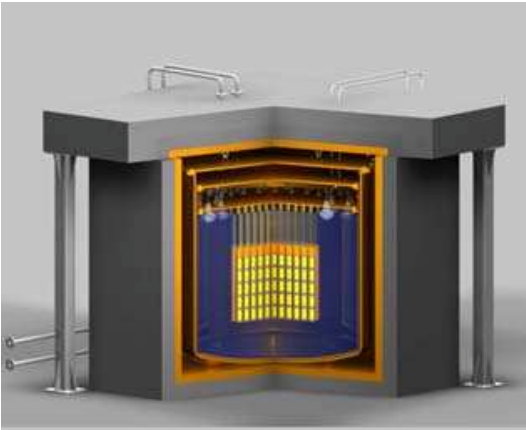}
\end{minipage}
\end{figure}
\section{Summary}

Astronomical observation, especially the cosmic microwave background
(CMB) radiation and the data on the large scale structure (LSS) of
the universe indicate that a significant part of matter content of
our universe is non-baryonic dark matter. The nature of the dark
matter is one of fundamental problems of particle physics and
cosmology. A favored candidate of the dark matter is the WIMPs,
weakly interacting massive particles. Direct searches of the WIMPs
are aiming at detecting the interaction of WIMPs with normal nuclei
which are SM particles. The CDEX (China Dark matter Experiment)
collaboration is a new  experimental group for searching dark matter
in the world whose task is to directly search for WIMP with an
Ultra-low energy threshold in terms of high purity Germanium
detector at CJPL (The China Jin-Ping deep underground laboratory).
So far, the CDEX collaboration includes several members: Tsinghua
university , Sichuan university , Nankai university, Institute of
Atomic Energy, Yalong River Basin Company and Nuctech Company.

The CJPL is located in the central portion of one of the transport
tunnels of a giant hydrodynamic engineering project at the huge
Jin-Ping Mountain area of Sichuan province,  southwest of China. The
rock covering thickness of CJPL is about 2400 m where the cosmic
muon flux is about 1 in $\mathrm{10^{-8}}$ of the ground level. The
Muon flux, radioactivity and radon concentration in the underground
lab have been measured and monitored time to time.

Comparing the detectors made of other materials the high purity
Germanium semi-conductor detector HPGe has many advantages such as
low radioactivity, high energy resolution, high density and
remarkable stability for radiation detection. The CDEX adopts the
high purity point contact Germanium detector PCGe with about 300 eV
threshold to search for WIMPs of minimum  mass as low as 10 GeV. The
detector mass of first phase CEDX1 is 1 Kg and that of second phase
CDEX10 is 10 kg.

There are two detectors of CDEX1, 20 g HPGe and 1 kg PCGe, running
now in the CJPL. The detector surrounded by CsI(Tl) or NaI(Tl) veto
detector and outside the veto detector there is a shielding system.
The performance of the detector has been calibrated and the noise
level is about 200 eV. The expectation of background count is less
than 0.1cpd/keV/kg near the 200 eV after event selection with the
PSD cut, noise edge cut and wavelet cut.

The CDEX-10 is a PCGe detector array which is immersed into a liquid
Argon ( LAr) vessel. Each unit detector in the array is a 1 kg PCGe
detector whose threshold is about 200 eV. The scintillation light in
the LAr will be read out by the PMT, and the LAr is the cooling
system providing working temperature for the Ge detector and also
serves as a veto detector. The Monte-Carlo study shows that the
background event rate will be as low as 1cpd at low energy range. In
the future, the CDEX Collaboration is going to set up a ton-scale Ge
detector composed of the PCGe detector and LAr active shielding and
cooling system in the CJPL. Hopefully,  the overall threshold of the
CDEX-1T detector will be less than 400 eV and the background event
rate could be reduced to 0.001 cpd. The sensitivity will be down to
about $\mathrm{10^{-45}cm^2}$ for the  WIMP with its minimum mass as
low as 10 GeV or even less.


\noindent{Acknowledgements}:\\

This work is partly supported by the National Natural Science
Foundation of China (NNSFC); Ministry of Science and Technology of China (MOSTC)
and Ministry of Education of China.\\

\end{document}